\documentclass[preprint,5p,times]{elsarticle}
\usepackage[T1]{fontenc}
\usepackage[utf8]{inputenc}
\usepackage{graphicx}
\usepackage{textcomp}
\usepackage{siunitx}
\usepackage{tabularx}
\usepackage{multicol}
\usepackage{lipsum}
\usepackage{verbatim}

\usepackage{ifthen}
\usepackage{eurosym}
\usepackage{fancyhdr}
\usepackage{lastpage}
\usepackage{rotating}

\usepackage{color}
\usepackage[finalnew]{trackchanges}

\journal{NIM A}

\begin{document}
\begin{frontmatter}
\title{RESPECT: Neutron Resonance Spin-Echo Spectrometer for Extreme Studies}
\author[e21,frm]{R. Georgii}
\ead{Robert.Georgii@frm2.tum.de}
\author[hopkins,e21]{J. Kindervater}
\author[e21]{C. Pfleiderer}
\author[e21]{P. B\"oni}
\address[e21]{Physik-Department, Technische Universit\"{a}t M\"{u}nchen,
  James-Franck-Str.\ 1, D-85748 Garching, Germany}
\address[frm]{Heinz Maier-Leibnitz Zentrum,
  Technische Universit\"{a}t M\"{u}nchen, Lichtenbergstr.\ 1, D-85748 Garching, Germany}
\address[hopkins]{Institute for Quantum Matter and Department of Physics and Astronomy, Johns Hopkins University, 3400 North Charles Street
Baltimore, MD 21218, USA}

\date{\today}

\begin{abstract}
    We propose the design of a \textbf{RE}sonance \textbf{SP}in-echo sp\textbf{EC}trometer for ex\textbf{T}reme studies, \textbf{RESPECT}, that is ideally suited for the exploration of non-dispersive processes such as diffusion, crystallisation, slow dynamics, tunneling processes, crystal electric field excitations, and spin fluctuations. It is a variant of the conventional neutron spin-echo technique (NSE) by i) replacing the long precession coils by pairs of longitudinal neutron spin-echo coils combined with RF-spin flippers and ii) by stabilizing the neutron polarization with small longitudinal guide fields that can in addition be used as field subtraction coils thus allowing to adjust the field integrals over a range of 8 orders of magnitude. Therefore, the dynamic range of RESPECT can in principle be varied over 8 orders of magnitude in time, if neutrons with the required energy are made available. Similarly as for existing NSE-spectrometers, spin echo times of up to approximately 1 microsecond can be reached if the divergence and the correction elements are properly adjusted. Thanks to the optional use of neutron guides and the fact that the currents for the correction coils are much smaller than in standard NSE, intensity gains of at least one order of magnitude are expected , making the concept of RESPECT also competitive for operation at medium flux neutron sources. RESPECT can also be operated in a MIEZE configuration allowing the investigation of relaxation processes in depolarizing environments as they occur when magnetic fields are applied at the sample position, i.e. for the investigation of the dynamics of flux lines in superconductors, magnetic fluctuations in ferromagnetic materials, and samples containing hydrogen. 

\end{abstract}
\end{frontmatter}

\section{Introduction}

\begin{figure}[htb]
\centering
\includegraphics[width=0.35\textwidth]{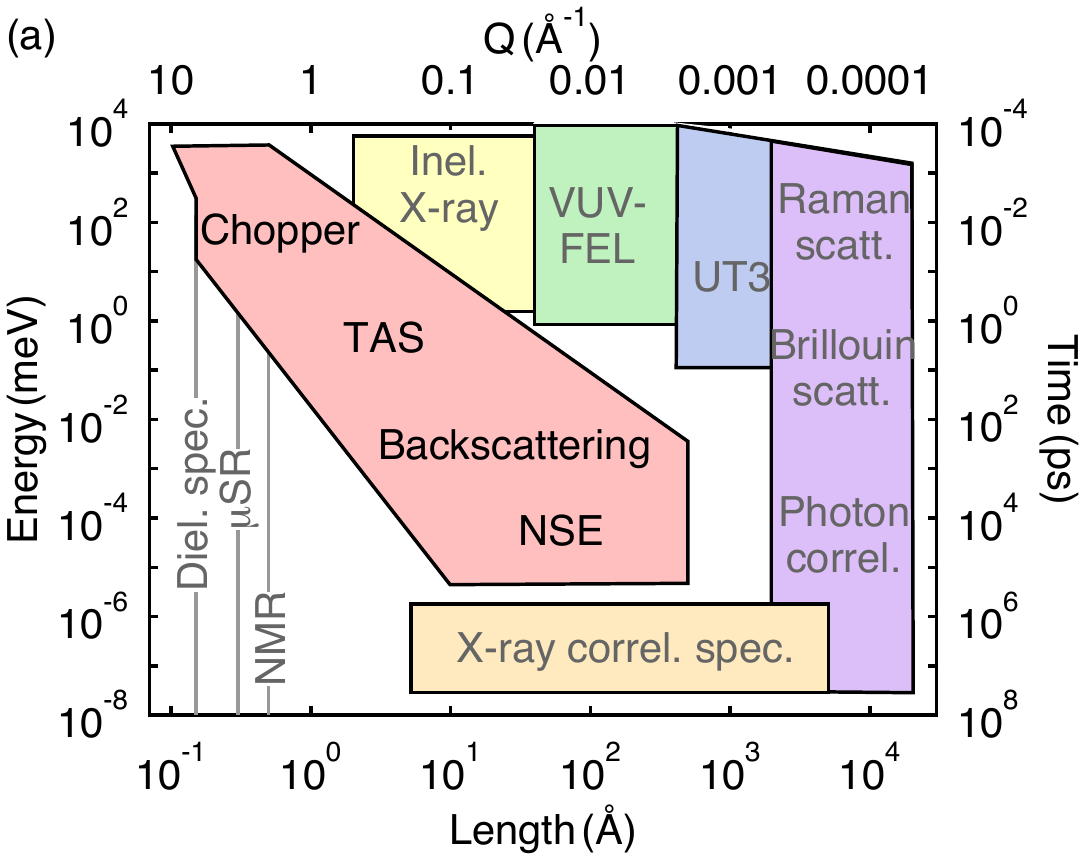}
\includegraphics[width=0.35\textwidth]{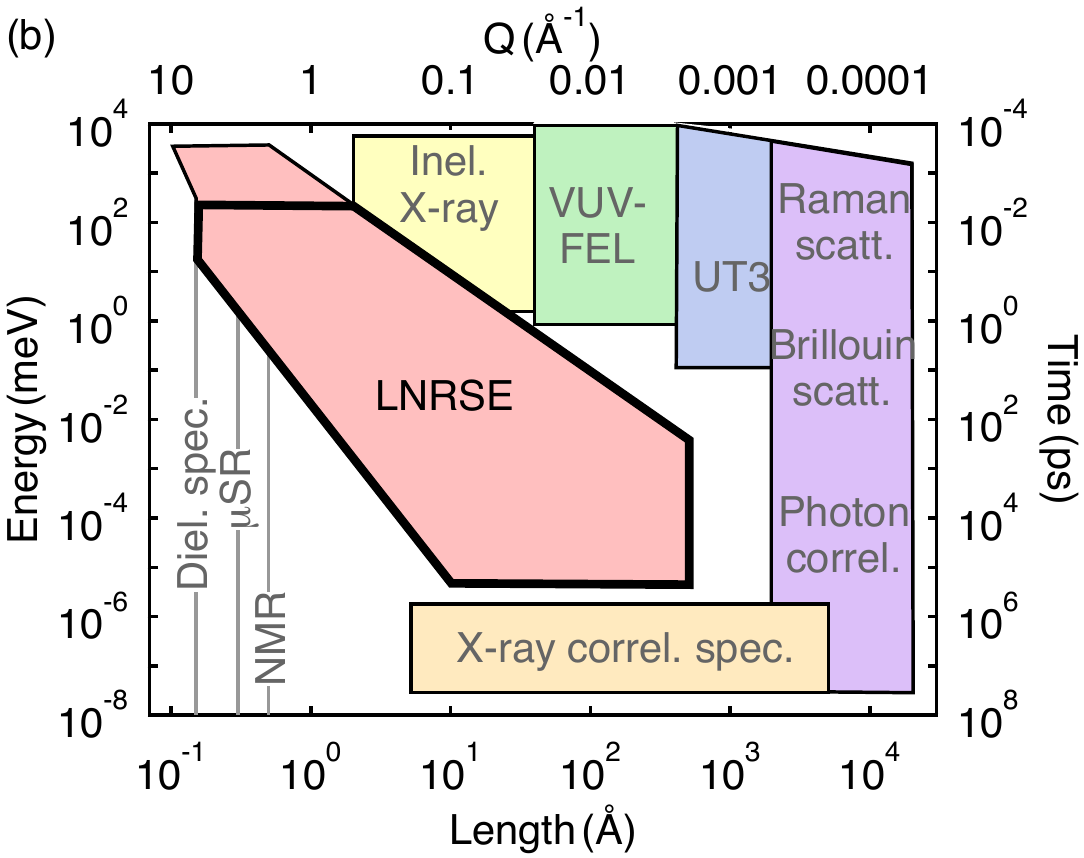}
\caption{\label{techniques}Dynamic range accessible with different microscopic probes in studies of soft and hard condensed matter systems. \add{For a better understanding both, the energy and the time axis are given. TAS resolutions are usually given in energy, for TOF and backscattering instruments both units are used and NSE dynamic ranges are typically expressed in time.} (a) Depiction of the regimes of conventional neutron scattering techniques. (b) Depiction of the regime of longitudinal neutron resonance spin-echo (LNRSE).}
\end{figure}

Neutrons are ideal probes to study static and dynamical properties of magnetic materials and systems containing light elements due to the large cross sections when compared with x-rays. Moreover, the possibility of contrast variation and the excellent energy resolution of typically 0.1 $\mu$eV to a few meV as achieved using conventional neutron scattering techniques such as triple-axis (TAS), time-of-flight (ToF), and backscattering spectroscopy (see Fig. \ref{techniques}) make the technique very attractive. With the increasing interest in studying slow processes at large spatial scales in the spin dynamics near quantum phase transitions and diffusive processes in soft matter systems it is important to develop instrumentation reaching time scales in the range of a few tens of ns towards 1 $\mu$s  \add{(i.e. in the neV range)}. Clearly, TAS or ToF are not suitable to achieve energy resolutions of this order because of the gigantic loss of intensity.

An elegant solution to circumvent the intensity problem is the use of neutron spin echo techniques, which allow a decoupling of the energy resolution of the instrument from the wavelength spread of the neutrons. The energy transfer of the neutrons is encoded in their polarization and not in the change of the wavelength of the scattered neutrons. Three variants of neutron spin echo spectrometers have so far been been realised: Neutron Spin Echo (NSE) \cite{Mezei:72}, Neutron Resonance Spin Echo (NRSE) \cite{Golub:87(M), Gaehler:88}, and the MIEZE technique in a transverse field geometry \cite{Besenboeck:98, Georgii:11}. Here, MIEZE is the abbreviation for ''Modulation of IntEnsity with Zero Effort''. The resolution and parameter ranges of these three types of spectrometers are limited due to various technical constraints.

In this paper we propose the design of a high-resolution NSE-spectrometer for a long pulse spallation source (LPSS) \cite{Mezei1995}
such as the European Spallation Source ESS or the currently proposed second target station at the SNS. Using the NRSE technique in a longitudinal field geometry, i.e. a  Longitudinal NRSE (LNRSE) similarly as first proposed by H\"au\ss ler et al. \cite{Haeussler2003501} we combine the advantages of both NSE and NRSE in one spectrometer. On the one hand, this combination allows to profit from the experience and knowledge of instrument design gathered over 40 years using NSE and NRSE as at ILL (IN15 \cite{IN15} and IN11 \cite{IN11}), at FRM-II (RESEDA \cite{RESEDA1} and JNSE \cite{JNSE}), at SNS (NSE \cite{Ohl:04}), at LLB (MUSES \cite{MUSES}) and at J-Parc (Vin Rose \cite{VinRose, Seto2016}). On the other hand LNRSE allows using the same correction techniques as the ones established in classical NSE and therefore has the potential to achieve at least the same energy resolution as classical NSE spectrometers. Because the essential difference between NSE and LNRSE is confined  to the Larmor precession regions only, the proposed concept for LNRSE is equally well adapted to continuous sources and short pulse spallation sources.

\begin{figure*}
\centering
\includegraphics[width=1.0\textwidth]{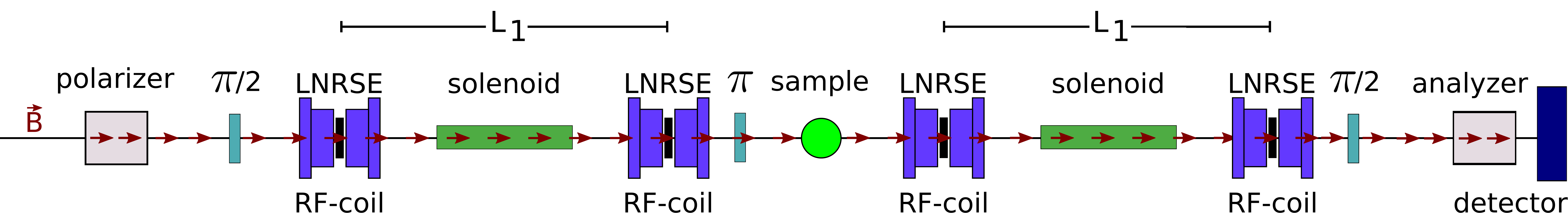}
\caption{\label{LRNSE1}Schematic depiction of the LNRSE configuration. \add{The static fields are printed in red. Note that the neutrons fly along the optical axis and  in contrast to transverse NRSE the orientation of the static ${\bf B}_0$-fields is parallel to the optical axis of the instrument, i.e ${\bf k_i} \parallel {\bf B_0}$. }The solenoids between the LNRSE coils provide the longitudinal guide field. In addition, they can be used to tune the spin echo times towards zero. Note that the length of the LNRSE coils ist typically of the order of a few tens of cm, i.e. much shorter than equivalent coils of length $L_1$ in a classical NSE instrument.}
\end{figure*}

The technique of resonant spin flips employed in LNRSE provides, in addition, the possibility to readily use the spectrometer in a MIEZE-1 configuration and thus to study samples under depolarising conditions as they occur in samples containing hydrogen or ferromagnetic domains and extreme sample environments such as high magnetic fields. Another optional upgrade using a MIEZE-2 configuration provides similar functionalities akin to the wide angle NSE-spectrometers SPAN at the former Hahn-Meitner Institute \cite{Pappas:2000} and WASP that is being realized at the ILL \cite{WASP}.


Let us summarize some of the key features of the LNRSE-technique as follows: i) The self-correction of the magnetic field inhomogeneites by the RF flipper coils reduces the energy density of the correction coils for diverging beams massively thus reducing also the small angle neutron scattering by the coils, i.e. for beams with zero divergence there is no need for correcting fields. ii) Due to the longitudinal guide fields no mu-metal shielding is required.  Moreover they can be used for field subtraction thus increasing the dynamic range of RESPECT by orders of magnitude. iii) The open design of the precession fields allows for the installation of focusing optics permitting gains in intensity of up to three orders of magnitude.

The paper is organised as follows: After providing a short introduction to the LNRSE technique and the two variants of the MIEZE-technique we explain the instrument concept of the {\bf RE}sonance {\bf SP}in-echo sp{\bf EC}trometer for ex{\bf T}reme studies (RESPECT). We discuss the phase space properties of the beam at the sample position and demonstrate that spin-echo times up to approximately 1 $\mu$s can be reached. In a next section we show in detail, how the guide system of RESPECT was optimized using the Monte-Carlo simulation package McStas \cite{Lefmann1999} and an analytical approach to determine the effects of divergence on the homogeneity of the field integrals. Finally, the results are summarized.

\section{Larmor Precession with Longitudinal Spin-Echo Coils}

This paragraph compares the salient properties of the LNRSE technique with those of the conventional NSE- and the NRSE-technique in a transverse field. In addition, the MIEZE-variants of LNRSE are introduced.

\subsection{Longitudinal Neutron Resonance Spin-Echo Spectroscopy}

In contrast to conventional NRSE \cite{Besenboeck:98, Georgii:11} the longitudinal setup is based on static fields ${\bf B}_0$ which are oriented parallel to the wave vector ${\bf k}_i$ of the incident neutrons (Fig. \ref{LRNSE1}). This geometry mimics the highly symmetric cylindrical field geometry of classical NSE. The radio frequency fields are installed perpendicular to ${\bf k}_i$ as in conventional NRSE. The precession region is defined by two $\pi/2$-flippers at the beginning and the end of the first and second spectrometer arm, respectively. 

Each arm contains two LNRSE coils. Along the flight path between the two $\pi/2$-flippers there is a longitudinal guide field provided by a solenoid to maintain the polarisation of the neutrons. If required, the precession can be reversed by a $\pi$-flipper in front of the sample. The static ${\bf B}_0$-coils consist of main and auxiliary coils in Helmholtz-configuration, where the auxiliary coils are producing an antiparallel field with respect to the main precession field ${\bf B}_0$. These auxiliary coils are used as in classical NSE to improve the field homogeneity at the entrance and exit of the coils and moreover to homogenise the magnetic field in the region of the radio frequency coils.

A dedicated LNRSE setup \cite{2014arXiv1406.0405K} has been realised at the instrument RESEDA \cite{RESEDA, RESEDA1} at MLZ (see also \cite{Kindervater:2015-a,Kindervater:2015-b}). \add{Using a beam size of 10 mm by 30 mm, reflecting the sample geometry and a vertical and horizontal divergence of $0.5^0$}, first test measurements demonstrate that up to an effective field integral of 0.160\,Tm no field corrections are required. \add {For larger field integrals the} usefulness of the standard correction coil technique was demonstrated in an earlier test experiment on IN11 \cite{Haeussler2003501, IN11}. In this experiment, the second conventional NSE arm was replaced by a LNRSE arm to compare directly the magnitude of the required correction fields, revealing that much smaller correction currents for the Fresnel coils are needed in the LNRSE arm. Therefore, much less material is in the beam and the demands for the cooling are tremendously reduced. In addition, small angle scattering is reduced too. 

As explained in paragraph \ref{field integrals}, the smaller correction currents are caused by two intrinsic properties of LNRSE: i) due to the spin flip in the center of the LNRSE coil most field inhomogeneities at the entrance of the coil are canceled by the similar field inhomogeneities at the exit of the coil. ii) the ${\bf B}_0$ coils are typically 10 times shorter than NSE coils reducing the field corrections accordingly. Moreover, the fabrication of short Helmholtz-coils is of minor technical challenge when compared with the fabrication of field-optimized coils for NSE.

\begin{figure*}[htb]
\centering
\includegraphics[width=1.0\textwidth]{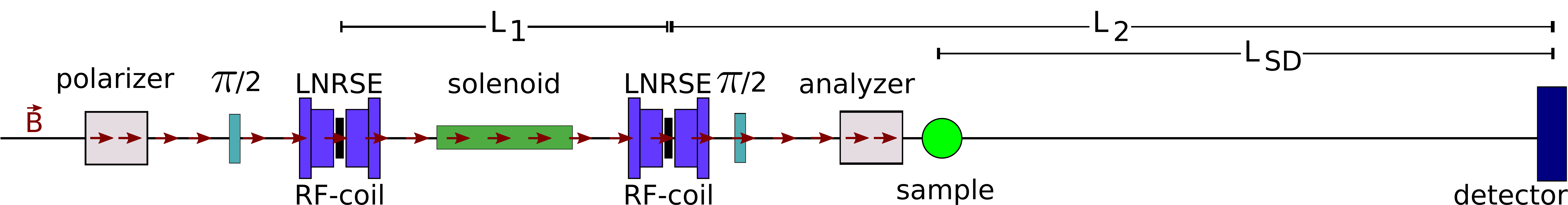}
\caption{\label{MIEZE}Schematic depiction of RESPECT in the MIEZE-1 configuration. It allows for investigating depolarising samples and samples in large magnetic fields in a small angle geometry. $L_2$ is the distance between the last RF-coil and the detector and is given by the MIEZE equation $L_2 = L_1 / \left( {\frac{\omega_2}{\omega_1} -1} \right)$ with $\omega_{1,2}$ being the two RF frequencies in the RF-coils. The MIEZE-time (equivalent to the NSE-time) is given by $\tau_M = \frac{h L_{SD} \omega_M}{m_n v^3}$, where $\omega_M = 2 \times (\omega_2 - \omega_1)$ and $v$ is the velocity of the neutrons.}
\end{figure*}

In contrast to the conventional transverse NRSE technique, the cylindrically symmetric LNRSE configuration allows guiding the polarization of the neutrons through the whole spectrometer and no spin rotations are required, reducing the effort to maintain the polarization. In addition, no bulky mu-metal shielding is required. Therefore, maintaining the polarization of neutrons with large wavelength $\lambda$ is facilitated.  These neutrons are particularly important as the resolution of the NSE-techniques increases with $\lambda^3$. 

The spin flip of the neutrons in the LNRSE coils allows subtracting efficiently field integrals by applying a magnetic field between the two LNRSE-coils by means of a solenoid (see Fig.  \ref{LRNSE1}) which provides a neutron guide field if run at low currents. The solenoid enhances the flexibility of the NRSE setup when compared with the NSE-technique as the spin echo times can be tuned continuously towards zero \cite{Haeussler2005} thus extending the range of accessible spin echo times by several orders of magnitude. Combining the high resolution of LNRSE with effective field integral subtraction allows covering nominally 8 orders of magnitude in one single set-up.

\subsection{Spectroscopy under Depolarising Conditions (MIEZE-1)}
A MIEZE-1 spectrometer requires only the primary arm of a (L)NRSE spectrometer (see Fig.~\ref{MIEZE}). MIEZE-1 has the advantage that all manipulations of the neutron spin are performed before the sample position. Therefore, depolarising samples and samples exposed to magnetic fields can be investigated. Even spherical polarization analysis is straight-forward because the beam leaving the MIEZE set-up is already polarized. Moreover, absorption and small angle neutron scattering are reduced and lead to a lower background when compared with LNRSE. However, because the MIEZE-1 method is sensitive to differences in the path length, measurements are restricted to small angels, small samples and/or samples with a special shape \cite{Georgii:11}. Therefore MIEZE-1 is most suitable in a SANS configuration.

MIEZE-1 is currently in user operation at two beam lines at the FRM II, namely RESEDA \cite{RESEDA1} and MIRA \cite{MIRA}. Its usefulness has been benchmarked in high magnetic fields up to 17 T \cite{2014arXiv1406.0405K} verifying that the contrast of the signal is maintained for MIEZE times $\tau_{M} = 15$ ns. Furthermore, $\tau_{M} = 105$ ns was reached at RESEDA  \cite{Kindervater:2015-b} using neutrons with $\lambda = 20$~\AA\
in zero magnetic field.  

\subsection{Spectroscopy over Wide-Angles (MIEZE-2)}
\begin{figure}[htb]
\centering
\includegraphics[width=0.45\textwidth]{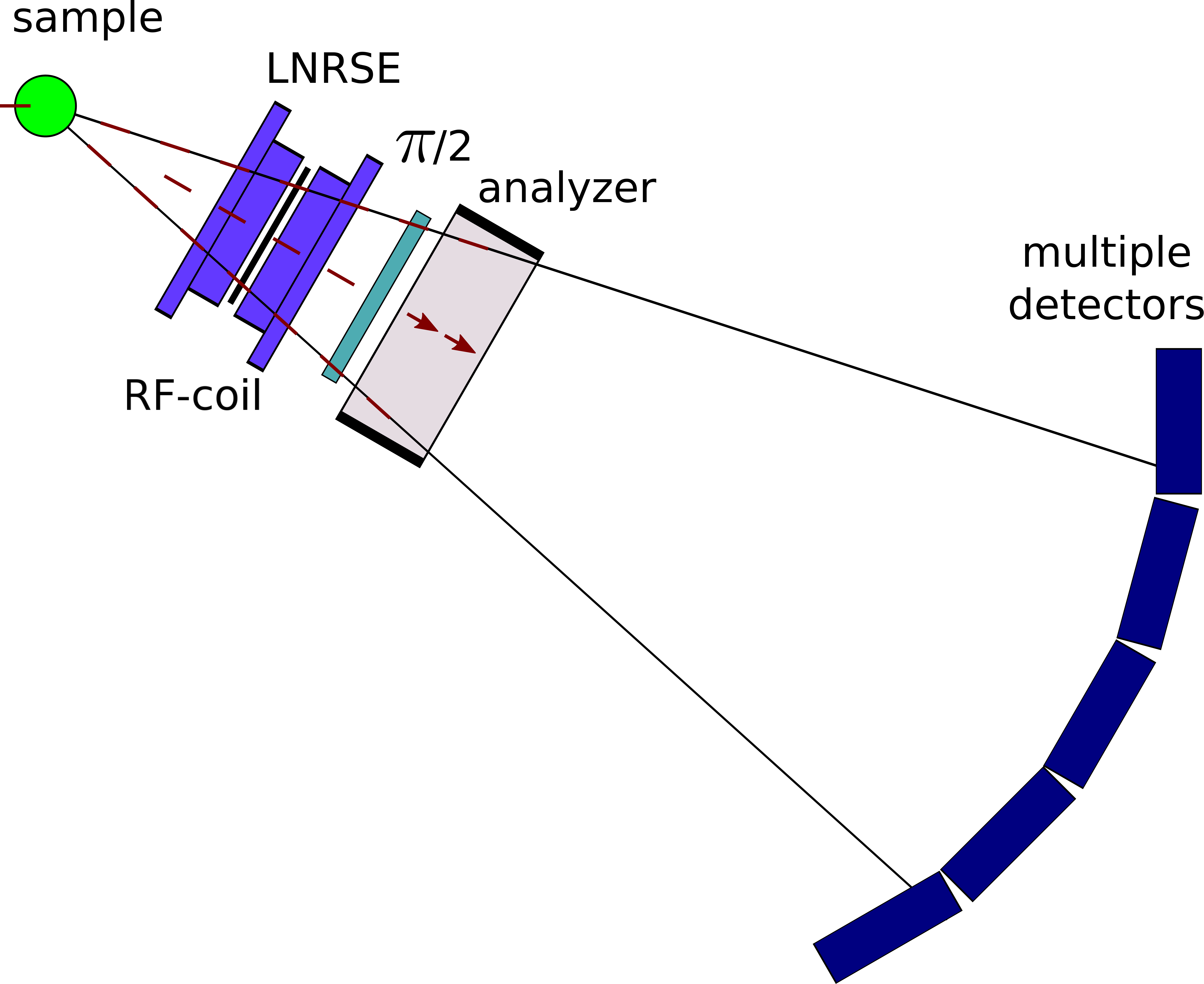}
\caption{\label{MIEZE-2}Schematic depiction of the secondary spectrometer of RESPECT in the MIEZE-2 configuration. MIEZE-2 allows for a measurement of the intermediate scattering function simultaneously over a wide $q$-range. The polarization has to be preserved at the sample position. The back precession is started at the wide-angle RF-coil behind the sample. In contrast to MIEZE-1, measurements under depolarizing conditions can only be performed with the addition of a polarizer. Differences in path length do not affect the measurements.}
\end{figure}

The restrictions of MIEZE-1 with respect to path length differences are eliminated by the installation of a third LNRSE coil after the sample (Fig.~\ref{MIEZE-2}). Here, no polarizer before the sample is required. After the sample, a wide angle RF-spin-flipper followed by a polarization analyzer is used in combination with a large area detector covering an angular range of $40^\circ$ or more. Larger solid angles can be covered by moving the detector closer to the sample. This technique is called MIEZE-2. However, in this case non-depolarizing samples must be used or an additional polarizer is required near the sample similarly as in standard NSE.

\section{Instrument Concept of RESPECT}

\begin{figure*}
\centering
\includegraphics[width=1.0\textwidth]{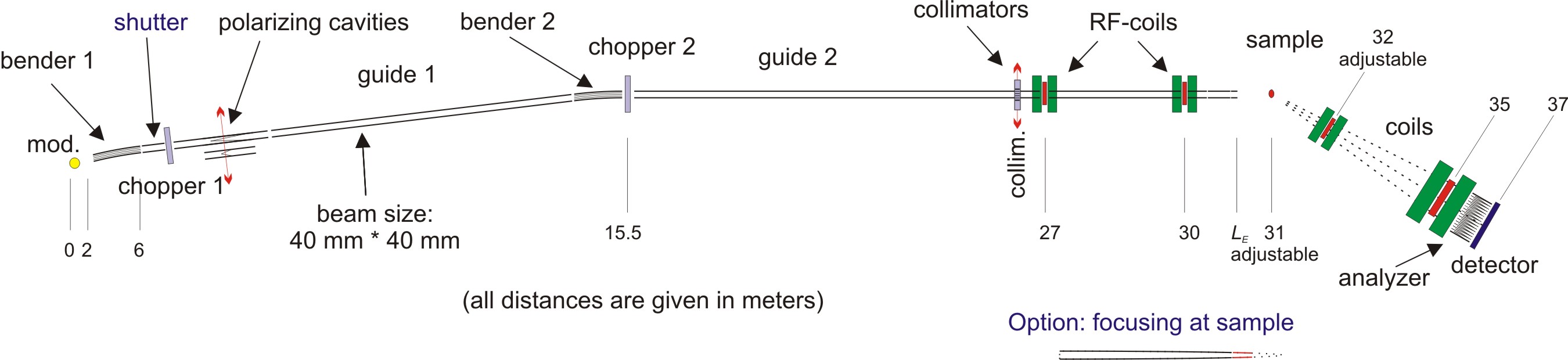}
\caption{\label{RESPECT_LNRSE}
Layout of the LNRSE spectrometer RESPECT. 
For the investigation of small samples with a volume of a few mm$^3$ focusing guides can be installed between the 30 meter position and the sample. A shutter is installed before chopper 1. The second bender is used to interrupt the line-of-sight from the moderator to the sample a second time. The two collimators (vertical / horizontal) before the first RF-coil are used to reduce the divergence of the beam if it is required for achieving high spin echo times and/or improved $Q$-resolution. \add{The second precession coil in the secondary arm will be enlarged to allow the use of an area detector.} For more details see text.}
\end{figure*}

In the following paragraphs we outline the design of the beamline RESPECT (Fig.~\ref{RESPECT_LNRSE}) optimised for a long pulse spallation source such as ESS that implements the LRNSE technique using state of the art technology for each component. The design goal is utilizing a wide wavelength band of approximately 2 \AA~ -- 22.25 \AA~ giving access to spin-echo times from less than 1 ps to more than 1 $\mu$s. For the calculations we assume as neutron source the pancake moderator of ESS \cite{ess_moderator}.

\subsection{Layout of the LNRSE Spectrometer}

\begin{figure}[htb]
\centering
\includegraphics[width= 0.4\textwidth]{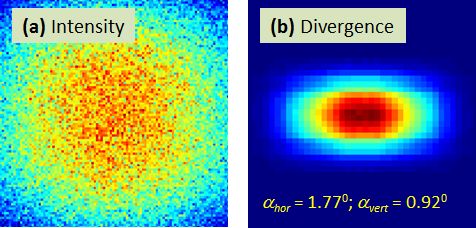}
\caption{(a) The beam at the sample position of RESPECT is  confined to the sample area if the guide ends at the second RF-coil. The position sensitive monitor displays an area 30 mm $\times$ 30 mm.  (b) The divergence is shown over an angular range of $4^0 \times 4^0$. The distribution is homogenous.  By inserting collimators upstream of the precession coils, the horizontal/vertical divergence $\alpha_i$ of RESPECT can be varied between $0.1^\circ/0.1^\circ$ and $1.8^\circ/0.9^\circ$.}
\label{Phase_Space_RESPECT}
\end{figure}

\begin{figure}[htb]
\centering
\includegraphics[width= 0.55\textwidth]{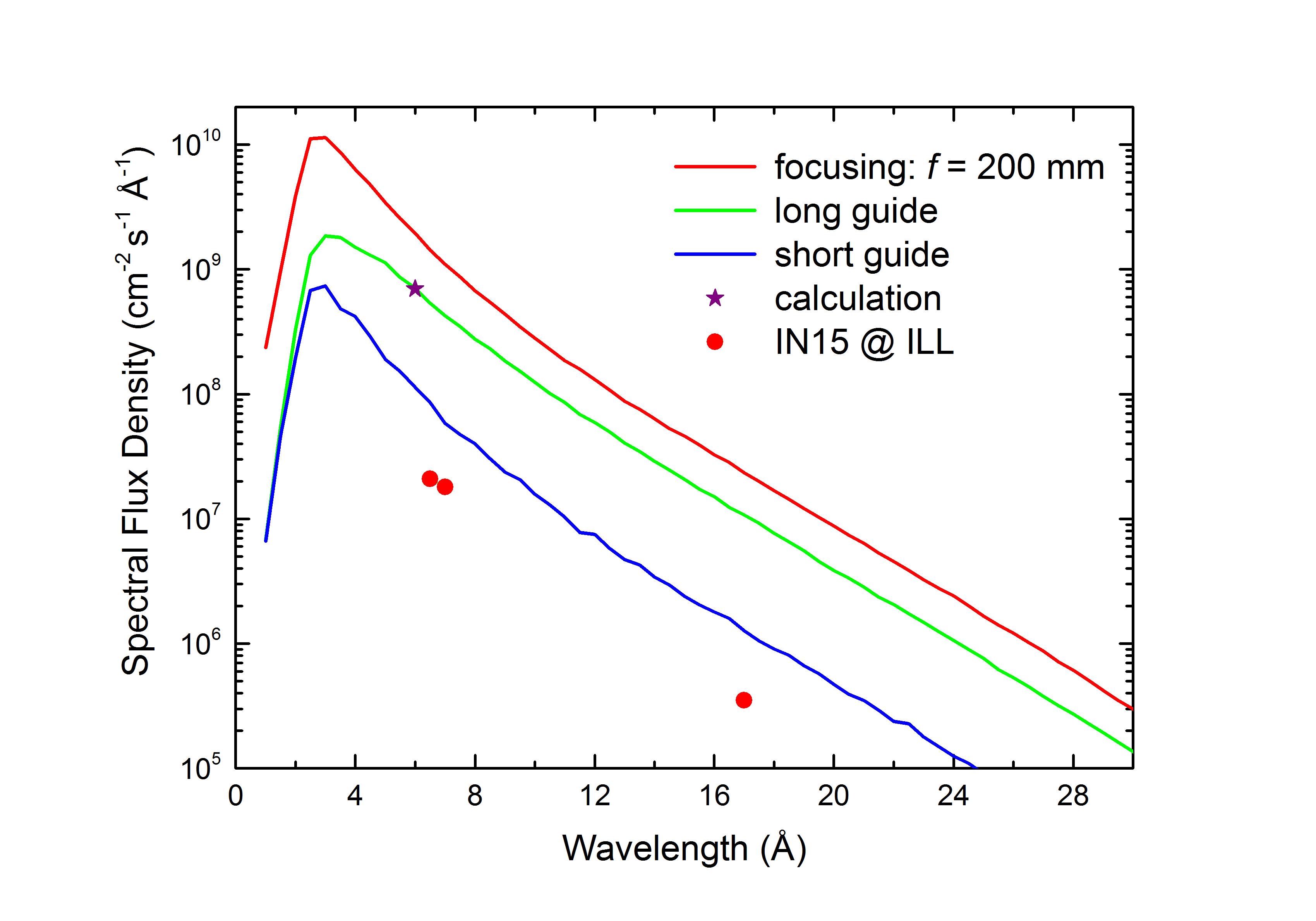}
\caption{The figure compares the spectral flux density at the sample position of RESPECT with the NSE-spectrometer IN15 \cite{ESSENSE}. The blue and green lines are simulations using McStas for a neutron guide ending before (short guide) and after (long guide) the RF-coils in the first arm, respectively. The red line gives the flux when a focusing guide is added to the set-up that ends 200 mm before the sample. The direct calculation of the spectral flux of RESPECT (purple star) at $\lambda = 6$ \AA\ is in perfect agreement with the McStas results.}
\label{Intensity_PanCake}
\end{figure}

The overall length of the instrument of approximately 37 m allows selecting three wavelength bands. To provide flexibility for bulky sample environment it is possible to increase the distances between the sample and the end of the primary and the beginning of the secondary precession regions when compared with the standard set-up.
The neutrons are transported using a neutron guide with a cross section 40 mm $\times$ 40 mm that is well adapted to the height of the pancake moderator ($h_{mod} = 30$ mm) and yields a homogenous illumination of the sample avoiding an illumination of the sample environment (Fig. \ref{Phase_Space_RESPECT}). The guide bridges the first precession region and ends at 30 m, i.e. 1 m before the sample (''long guide'' in Fig. \ref{Intensity_PanCake}). The Monte-Carlo simulations were performed using the flat spectrum provided by the software program McStas  (Versions 1.12c and 2.1 \cite{Lefmann1999, McStas} and the component file: source\_gen.comp \cite{McStas}). For more details see \ref{BeamProperties}.

The spectral flux density of RESPECT is compared with that of the existing NSE beamline IN15 at the ILL in Fig.~\ref{Intensity_PanCake}. The simulations (solid lines) show that RESPECT yields a significantly higher spectral flux density at the sample in particular at short wavelengths. For the simulations with McStas the component for the ESS pancake moderator was used \cite{ess_moderator}. The simulation for the long guide is in excellent agreement with the calculatation of the spectral flux density at $\lambda = 6$ \AA\ (purple star), which is based solely on i) the brilliance of the pancake moderator, ii) phase space considerations, and iii) the transmission functions of the various components (see paragraph \ref{Mean_Brilliance}). Inserting a parabolic focusing guide between the first RF-coil and the sample leads to a beamsize of approximately 10 mm $\times$ 10 mm and a spectral flux density exceeding $1\cdot 10^{10}$ cm$^{-2}$s$^{-1}$\AA$^{-1}$.

\subsection{Primary Spectrometer}
In the following the essential components of the guide system are discussed in detail following downstream along the beamline from the moderator to the sample.

\subsubsection{Chopper System}
The LNRSE spectrometer RESPECT will be mostly used for investigations of dynamic processes in soft matter and in magnetism at the very long time scales available for wavelengths of the order of 20 \AA\ and at shorter time scales by taking advantage of the flux maximum of the pancake moderator around 3 \AA. The first decision to be made concerns the selection of the width of the wavelength bands to be used. One frame with a bandwidth of $\triangle \lambda \simeq 20$~\AA~ is not feasible because the instrument length given by
\begin{equation}
	L = \Delta T/(\alpha \Delta\lambda)
\end{equation}
would have to be restricted to $\simeq 14$ m in order to use each pulse provoking serious problems with background and space to operate RESPECT. Here, $\Delta T = 1/f$ is the time between the pulses (at ESS: $f = 14$ Hz), $\alpha = 252\, \mu$s/\AA/m is a constant and $\Delta \lambda$ is the wavelength band. Similarly, two frames would yield a too short instrument creating serious background problems. Using more than three frames would decrease the integrated intensity over the anticipated wavelength bands and hamper the operation of RESPECT at large wavelengths where the brilliance of the pancake moderator is small. 

\begin{figure}[htb]
\begin{center}
\includegraphics[scale=0.5]{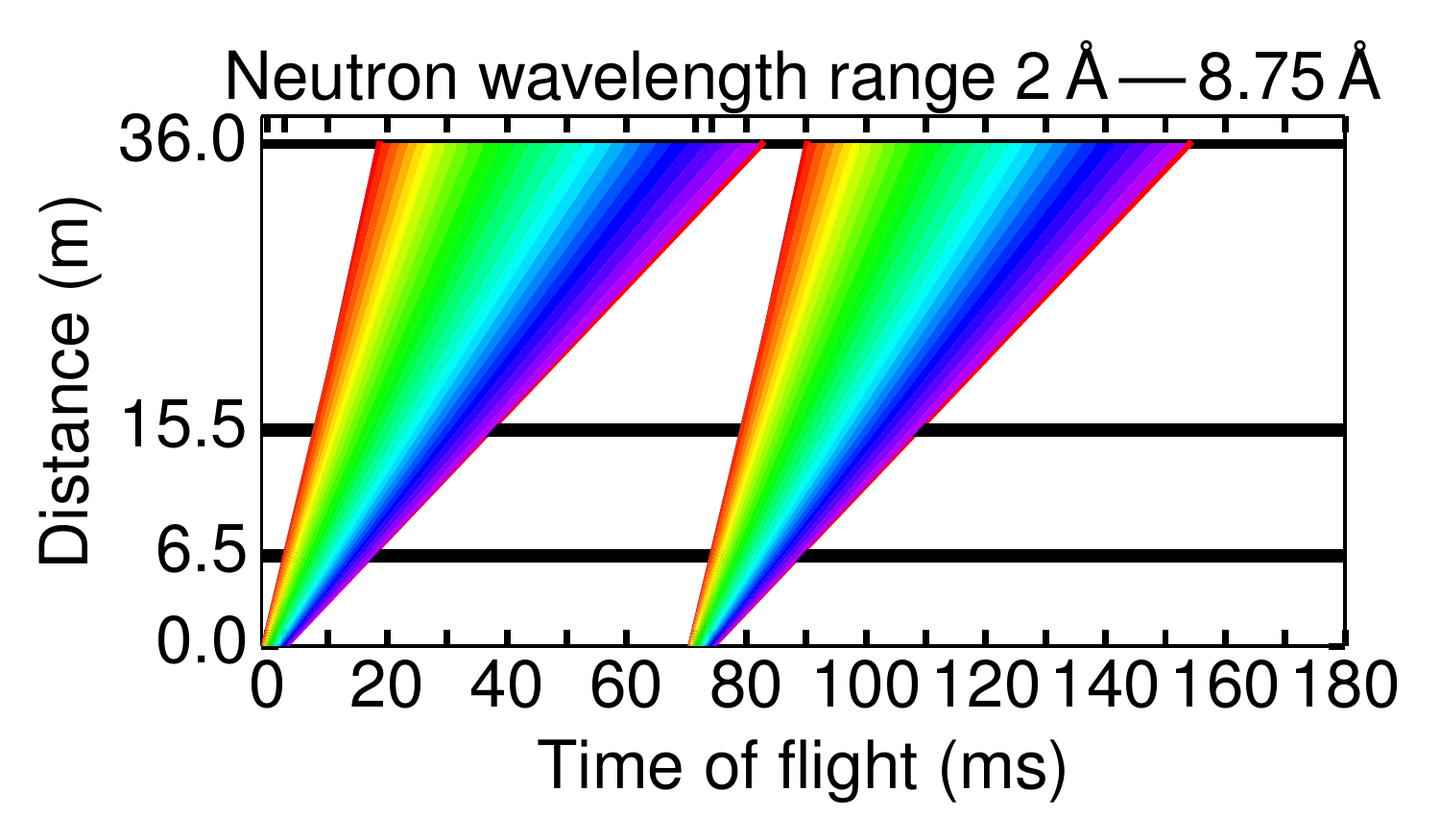}
\includegraphics[scale=0.5]{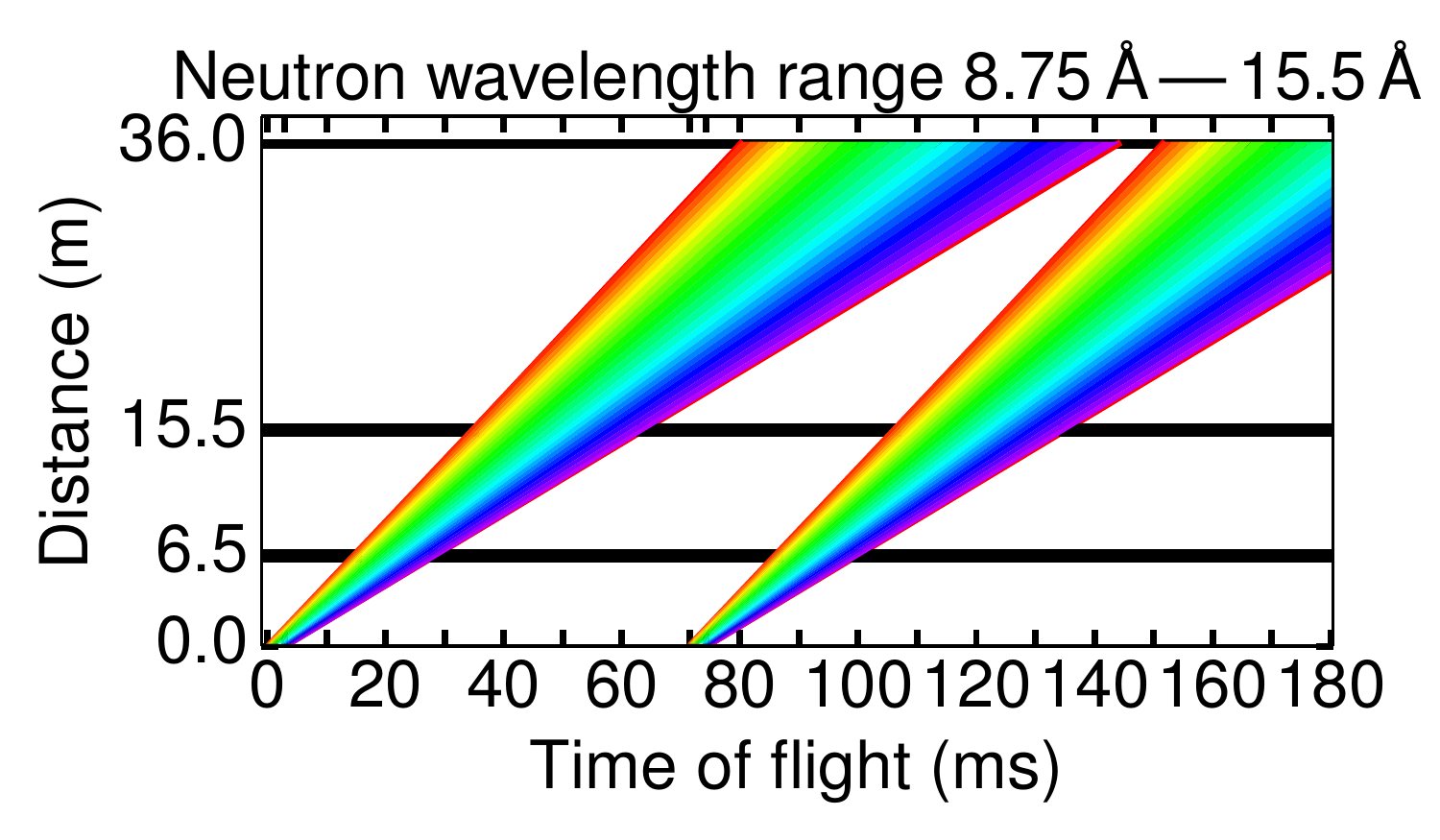}
\includegraphics[scale=0.5]{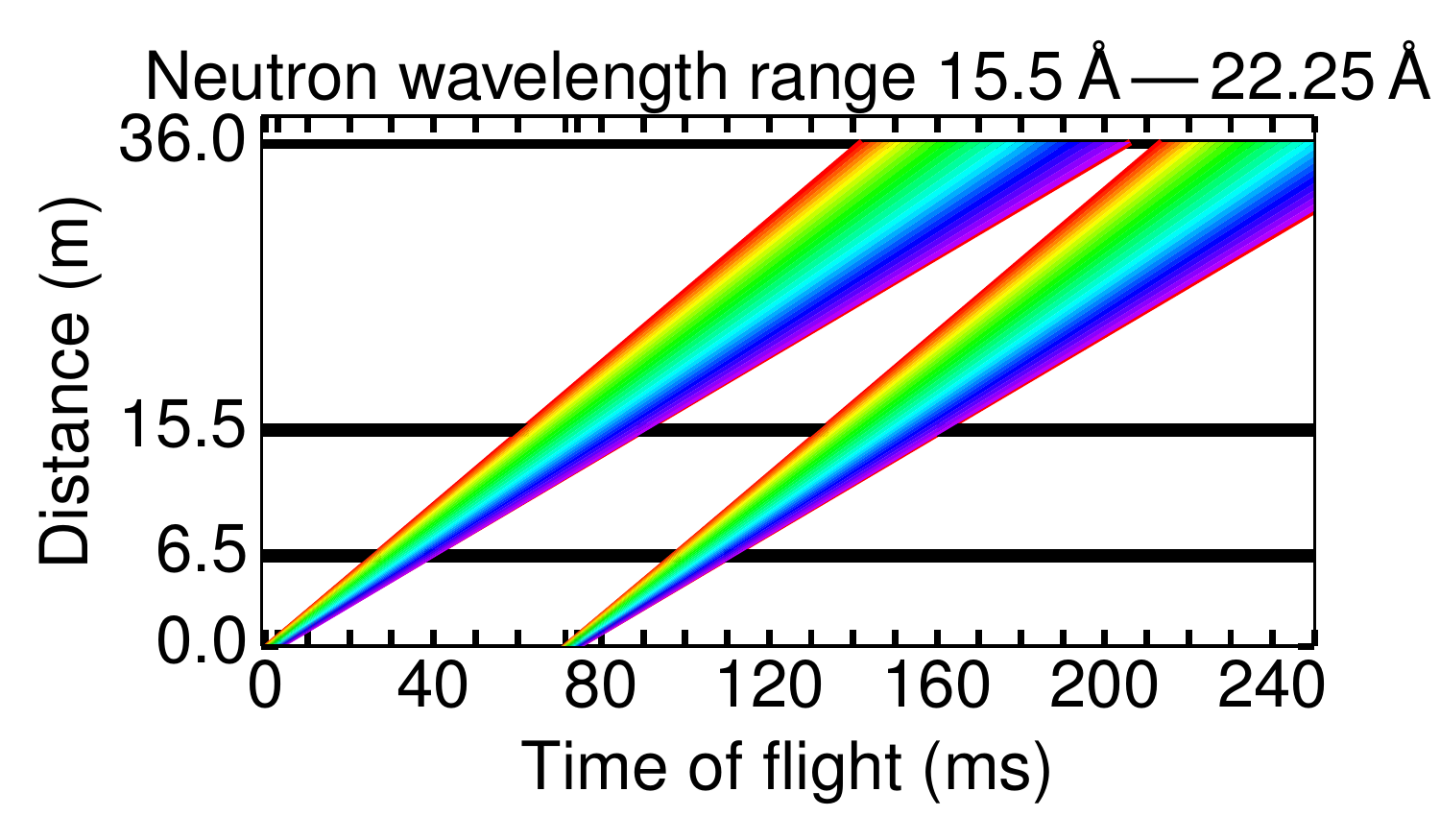}
\end{center}
\caption{\label{Chopper}The three wavelength bands, which are proposed for RESPECT. The opening times for the wavelength frame selection choppers 1 and 2 are 11.41 ms and 34.03 ms, respectively.}
\end{figure}

Therefore, we propose using three frames leading to bandwidths of approximately 6.75~\AA~thus covering a wavelength range from 2~\AA~to 22.5~\AA\ as shown in Fig. \ref{Chopper}. Using standard chopper systems and respecting a safety margin between frames leads to an instrument length of 37 m. The three wavelength bands are defined by choppers 1 and 2, which are adjustable by their phase shift. They are positioned 6.5 m and 15.5 m downstream of the moderator (Fig.~\ref{RESPECT_LNRSE}). In principle, the pulse is defined by the time structure of the moderator, however, we propose the installation of an additional chopper for pulse shaping to define a well-defined beam by cutting the tails of the ESS pulse.

Based on the above configuration, the wavelength resolution is given by
\begin{equation}
\frac{\delta \lambda}{\lambda} = \frac{\tau}{\alpha L \lambda} \; \mbox{,}
\end{equation}
where $\tau = 2.86$ ms is the length of the neutron pulse. One obtains ${\delta \lambda / \lambda} = 5\%$ at 6 \AA\  and 1.6\% at 16 \AA, which are reasonable values for applications as anticipated. The wide frame at large wavelengths compensates for the intensity losses due to the decreasing flux at large $\lambda$. Furthermore due to the small beam size and the compact chopper system thanks to the slim guide the pulses will be chopped in a very clean way.  There is also no need for a frame overlap chopper, thus nearly the full intensity per pulse can be delivered to the sample.

\subsubsection{Bender 1}

The bender comprises 4 channels thus providing a homogenous intensity across the beam as opposed to a curved guide. It has a length $L_{bend} =  1.2L_S = 3.96$~m and a radius of curvature $R_b = 140.34$ m. $L_S$ designates the line of sight of an individual channel of the bender. The sides are coated with supermirror $m = 4.5$ leading to a deflection angle $\psi = 1.6^\circ$ and a low critical wavelength $\lambda^* = 1.51$ \AA. The small $\lambda^*$ leads to an excellent transmission for $\lambda \ge 2$ \AA, \change{which is matched very well to the down-stream guide system that uses coatings $m = 2$.}{which is matched very well to the wavelength dependence of the brilliance of the pancake moderator of ESS  (see Fig. 13 below) and the down-stream guide system that uses coatings m = 2.} Of course, by increasing the number of channels and $m$, the transmission at small $\lambda$ could and actually should be improved further. The effective line-of-sight is interrupted a couple of meters away from the biological shielding, i.e. much quicker than by using a curved guide.

The use of metallic substrates (made from Cu or Al) \cite{boeni2010} for the body of the bender and Si-wafers for the blades will reduce the flux of high energy neutrons and $\gamma$ radiation, leading to cost savings in shielding. The exit of the bender may be equipped with a thin Al window to decouple the vacuum (or $^4$He environment) of the bender from the first chopper and the straight guides. 

\subsubsection{Guide System}

The proposed guide system with a cross section 40 mm $\times$ 40 mm  (Fig.~\ref{RESPECT_LNRSE}) is well adapted to a sample size of  30 mm $\times$ 30 mm and the height of the moderator $h_{mod} = 30$ mm, yielding a maximum vertical and horizontal divergence of $0.92^\circ$ and $1.8^\circ$, respectively (Fig. \ref{Phase_Space_RESPECT}).
The compact phase space of the neutrons allows not only for high-resolution experiments but also experiments requiring a high intensity at a reduced resolution, i.e. short spin-echo times as obtained for short $\lambda$ near the flux maximum of the moderator. Moreover, due to the small cross section of the guide, effects of inhomogeneities of the field integral and stray fields from the environment on the Larmor precessions are reduced. For more details see paragraph \ref{BeamProperties}.

\subsubsection{Polarizing cavities}

Polarizing cavities are an efficient means to polarize neutrons over a wide range of wavelengths \cite{boeni2009}. However, if the taper angle $\epsilon$ of the polarizing blades becomes comparable to the critical angle of reflection of the neutrons for the spin down neutrons ($m \simeq 0.68$) also the properly polarized neutrons are reflected out from the beam. On the one hand, these are the most valuable neutrons because they have a small divergence. On the other hand, cavities remove neutrons with very long wavelengths therefore a filter to remove these unwanted neutrons is not required. Hence, at least two cavities with $\epsilon = 0.35^\circ$ and $1.9^\circ$ for the wavelength ranges 2 \AA\ $\le \lambda \le 10$ \AA\ and 8 \AA\ $\le \lambda$, respectively, should be installed.  The polarizing coatings are made from FeSi-supermirror with $m = 4$.

\begin{figure}[htb]
\centering
\includegraphics[width= 0.5\textwidth]{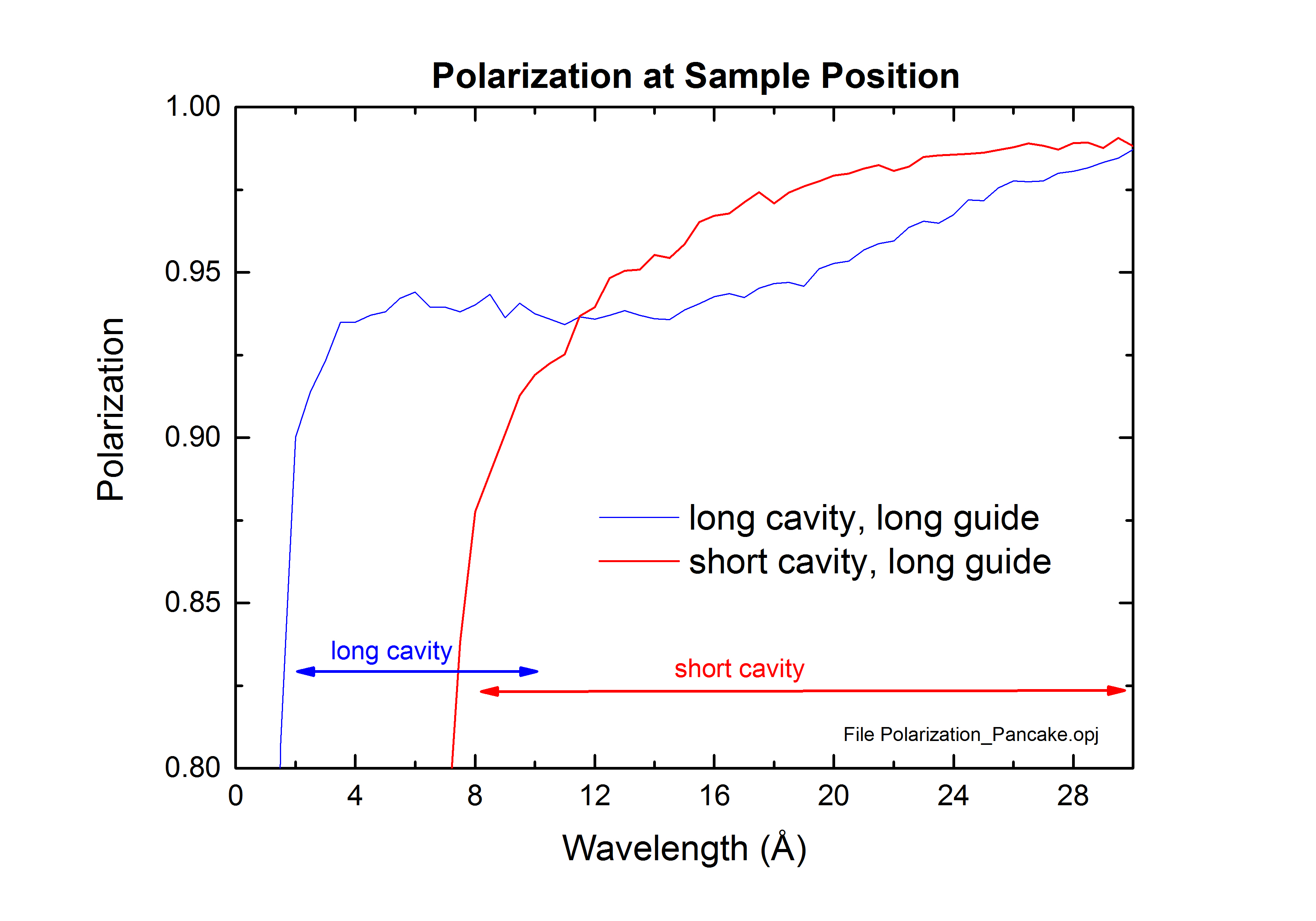}
\caption{The figure shows the polarization of the short and the long cavity for small and large wavelength ranges, respectively. Effects of multireflections of the neutrons within the Si-wafers, which are coated on both sides, and the absorption of the neutrons by the Si wfers and the Fe/Si coatings are taken into account.}
\label{PolarPancakelong}
\end{figure}

Fig.~\ref{PolarPancakelong} shows the polarization of the transmitted neutrons as calculated using the software package McStas. A reasonably high polarization between $92\%$ and $ 98\%$ is obtained. To achieve polarizations $P > 99\%$ double V-cavites \cite{boeni2009}
may be installed. Note that for simulating the flux of the polarized neutrons at the sample position (green solid line in Fig. \ref{Intensity_PanCake}), the long cavity was used for the complete range of wavelengths 2 \AA\ $\le \lambda \le 30$ \AA. Therefore, the flux at large $\lambda$ is underestimated.

\subsubsection{Bender 2}

Bender 2 has the same geometry as bender 1. It guarantees that all neutrons are reflected at least two times before reaching the sample to reduce background. In contrast to bender 1, the guide body and the dividing blades are manufactured from glass.

\subsubsection{Collimation stage}
\label{divergence-max}

\begin{figure}[htb]
\centering
\includegraphics[width=0.35\textwidth]{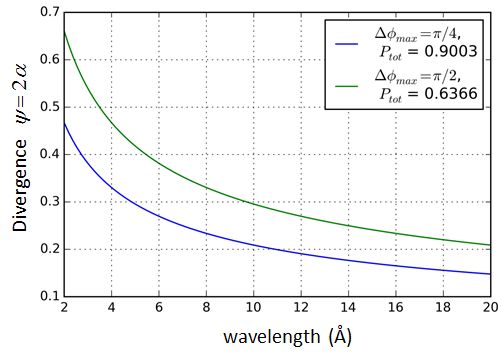}
\caption{The divergence versus wavelength for $P_{tot} = 90\%$ and 64\% are shown as a blue and a green line, respectively. For details of the calculations see \ref{field integrals}. The field integral is 1.02 Tm.}
\label{3A_94_Div}
\end{figure}

The vertical and horizontal divergence of the neutron beam at the sample position of RESPECT without collimation is $0.92^\circ$ and $1.77^\circ$, respectively (Fig. \ref{Phase_Space_RESPECT}). As shown in paragraph \ref{field integrals}, a large divergence may be detrimental for the resolution of Larmor precession techniques if no correction coils are used (Fig. \ref{3A_94_Div}). However, the essential feature of RESPECT is  that for a beam with a divergence as large as $\psi = 0.2^0$ field integrals $J =1.02$ Tm can be realized at a polarization of 64\%. The expression
\begin{equation}
	\tau = {2\pi m_n^2\over \hbar^2}\gamma J \lambda^3
\label{tnse}
\end{equation}
for the spin-echo time can be written in a simplified form
\begin{equation}
	\tau{\rm[ns]} = 0.186J{\rm [Tm]}(\lambda{\rm [\AA]})^3
\end{equation}
when the natural constants are inserted. Here, $m_n$ is the mass of the neutron and $\gamma$ its gyromagnetic ratio. For example, for the wavelengths $\lambda = 3$ \AA\ and 20 \AA\ one obtains nominally $\tau= 5.1$ ns and 1.5 $\mu$s, respectively.

To define the divergence of the beam at the sample position, various vertical and horizontal collimations between 10 min and 60 min can be driven into the beam \cite{Komarek2011}. Obviously, the use of tight collimations will supersede the effects of the downstream neutron guide on the beam divergence and the intensity. If focusing guides are installed between the exit of the precession region near the 30 m and the $L_E$ positions, the reduction of the divergence of the beam will lead to an improved focusing of the beam at the sample position. Experience shows that a beam size of the order of a few mm can be achieved.

\subsubsection{Focusing guides}

For measurements on samples with a size smaller than 30 mm $\times$ 30 mm, parabolical focusing devices will be introduced bridging the flight path between the last precession coil and the position $L_E$ (Fig. \ref{RESPECT_LNRSE}), which may be as close as 80 mm from the sample (see for example \cite{adam2014, VINROSEGuides}).

According to Liouville's theorem, the flux density at the sample position increases roughly inversely proportional to the area of the beam. Therefore, reducing the beam size from 30 mm $\times$ 30 mm to 10 mm $\times$ 10 mm yields a gain of almost a factor of 10. If the resolution conditions allow, one may end up (for example at $\lambda = 6$ \AA) with a spectral flux density at the sample exceeding $1\cdot 10^{10}$ cm$^{-2}$s$^{-1}$\AA$^{-1}$ polarized neutrons (Fig. \ref{Intensity_PanCake}) indeed a very intense beam \cite{Hils, Kardjilov}. 

\subsubsection{LNRSE coil system}
\label{coilsystem}

The RESPECT spectrometer will include four LNRSE coils. Two of them in the primary spectrometer arm. The static ${\bf B}_0$ coils consist of normal conducting, water cooled windings designed to produce a magnetic field $B_0 = 0.17$\,T. With a coil distance of $L_1=3$\,m, employing $\pi$-flips, the effective field integral of one spectrometer arm will be $J = 0.17$\,T $\cdot$ 3\,m $\cdot\,2 = 1.02$~Tm yielding according to Eq. (\ref{tnse}) nominal spin echo times $\tau=190$\,ns and $\tau=1.5\,\mu s$ at $\lambda = 10$~\AA~and $\lambda =20 $~\AA, respectively. The radio frequency coils will be operated at $\approx 5$\,MHz. The $q$-range in the LNRSE configuration will extend to at least 4.5\,\AA$^{-1}$. For details concerning field integrals in LNRSE see \ref{field integrals}.

\subsection{Secondary Spectrometer}

The secondary spectrometer will be very similar to a classical NSE-spectrometer the major difference being the replacement of the solenoids for NSE by two RF-coils for LNRSE. In the following we make a few remarks concerning the layout of the components.

\subsubsection{LNRSE coil system}

The coil system for the secondary arm is almost identical to the system in the primary arm. To allow for a large detector with dimensions of approximately 30 cm $\times$ 30 cm the RF- and $B_0$ coils are enlarged. This applies in particular for the second RF-coil as indicated in Fig. \ref{RESPECT_LNRSE}.

\subsubsection{Analyzer}

A large area polarizing analyzer as used for example for the beam line JNSE at FRM II \cite{JNSE} will be installed. The blades have the dimensions of typically 30 cm $\times$ 30 cm and are coated with remanent supermirror FeCoV/TiN$_x$ \cite{boeni1996, boeni1999}.
However, as the reflectivity of Fe/Si supermirrors is very high and because LNRSE is less sensitive to magnetic fields, more advanced coatings such as Fe/Si with $m = 5.5$, a reflectivity of 70\% at $m = 5.5$, and a polarization $P > 99\%$ 
will be considered \cite{schanzer2016}. The newest generation of coatings can be magnetised in fields as small as 20 mT. To respect the symmetry of LNRSE, the coatings will be magnetized in a longitudinal field.

\subsubsection{Detector}
\add{A similar detector array as used for the existing NSE-spectrometers will be considered for the NRSE part of RESPECT.}
These detectors cover typically an area of 30 cm $\times$ 30 cm and provide a moderate (2D) position resolution (better than 3 cm), high efficiency and high maximum count rate (better than 5 kHz/cm$^2$). A robust solution yielding a resolution of approximately 8 mm would consist of an array of $\simeq 40$ front end counting tubes ($^3$He) that are read out using charge division.

\subsection{Further Instrument Add-on Options}
\add{Removing or replacing the second LNRSE arm together with using a separate detector system yields two possible MIEZE options as explained in the following.}

\subsubsection{MIEZE-1}
\label{mieze-1}

\begin{figure}[htb]
\centering
\includegraphics[width=0.42\textwidth]{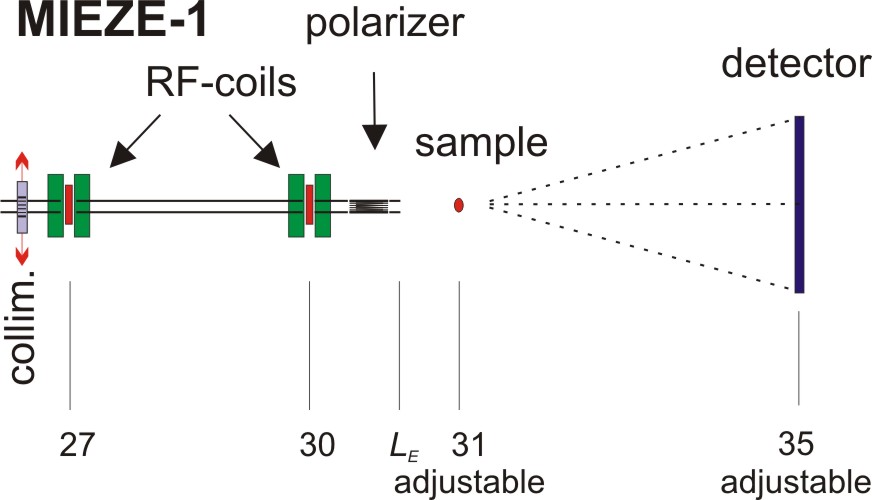}
\caption{\label{RESPECT_MIEZE-1}
For the longitudinal MIEZE option, the secondary precession coils are removed and a large detector with an excellent time resolution is placed in the forward direction. In addition, a short polarizing transmission bender is inserted between the last RF-coil in the primary arm and the sample.}
\end{figure}

For MIEZE-1 a polarizing transmission bender is inserted before the sample  (see Fig.~\ref{RESPECT_MIEZE-1}). The direction of its magnetising field is along the beam direction (longitudinal) to respect the symmetry of the field configuration of the beam line. Compensation fields will be installed to reduce the disturbance of the precession fields by the magnetising fields of the polarizers. 

A fast detector system is needed to detect the high frequency intensity variations of the MIEZE signal. It will consist of an array of  32 CASCADE detectors ($25\times25\,{\rm cm^2}$) for the small angle MIEZE setup.  The CASCADE detector concept of CDT \cite{Klein:2011jj} is a self-sufficient system comprised of a detector front end and an integrated detector readout system with on board histogramming electronics. The front end consists of 10 layers of GEM foils with $^{10}$B coating. A mixture of 85~\%Ar and 15~\%Co gas is used as counting gas and electron getter for avoiding sparking, respectively. Each of the foils allows reading out 128 by 128 pixels simultaneously. The signals of the different foils are added up in order to reach a detection efficiency of 50--60~\% for a neutron wavelength of 5~\AA. In the MIEZE mode, the 128 by 128 pixels of the ten foils can be counted in a histogram in time for each pixel separately. Together with the phase information from the radio frequency (RF) current this allows the detection of the MIEZE modulation in each pixel \cite{Haussler:2011ei}.

CASCADE detectors have been successfully commissioned at the instruments RESEDA \cite{Haussler:2011ei} and MIRA \cite{MIRA} at the FRM II. The correct working parameters together with the optimum mixing parameters for the detector gas were determined and are together with the resonant circuits in user operation.

The operation of a MIEZE-1 spectrometer using a pulsed neutron beam has already been tested recently \cite{Brandl:12} and the necessary adaption in the sweeping of the amplitude of the RF-fields has already been developed. It is planned to base the data evaluation on the existing ILL code library for NSE instruments as used at the ILL and the FRM II. This code already includes the necessary adaptation for pulsed beams. A recently developed software package \cite{ISI:000340245000034} for graphical data evaluation based on the ILL library could be adopted to the ESS software suite.

\subsubsection{MIEZE-2}
\label{mieze-2}
\begin{figure}[htb]
\centering
\includegraphics[width= 0.35\textwidth]{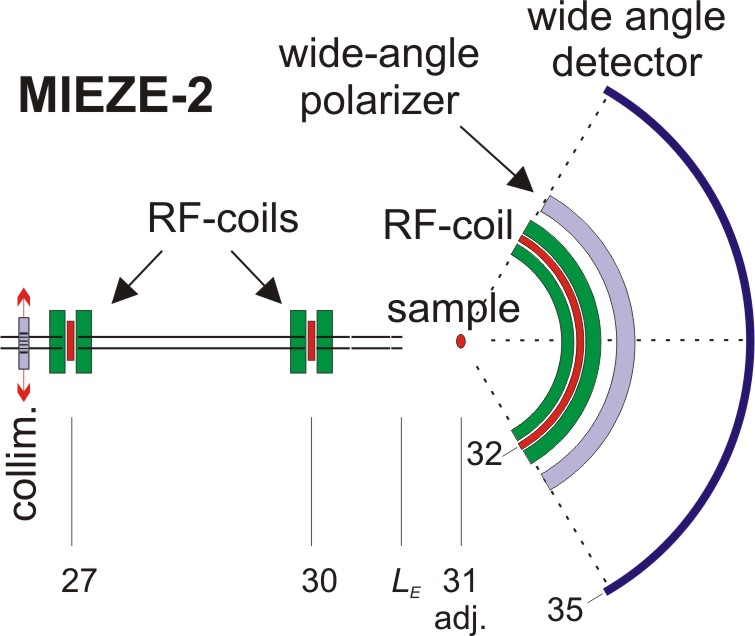}
\caption{\label{RESPECT_MIEZE-2}
The longitudinal MIEZE-2 option is realized by replacing the secondary LNRSE-arm by a wide-angle RF-coil followed by a wide-angle analyzer. A time-resolved wide-angle detector, for example CASCADE, monitors the time-dependent signal. This set-up is ideally suited for the investigation of non-depolarizing samples. For the investigation of small samples, parabolic focusing elements can be installed in the primary arm.}
\end{figure}

For experiments at large scattering angles and if path-length differences become an issue \cite{Brandl:11},  the MIEZE-2 configuration will be used (see Fig.\ref{RESPECT_MIEZE-2} ).  Here, no polarizer before the sample is required. After the sample, a wide angle RF-spin-flipper followed by a polarizer will be used covering an angular range of approximately $40^\circ$. A similar CASCADE system as for MIEZE-1 can be used. As the system needs to cover a much larger solid angle, 32 additional detectors for covering the whole solid angle provided by the wide angle coil are required. Larger solid angles can be covered by moving the detector closer to the sample.

MIEZE-2 will provide similar functionalities as the wide angle NSE-spectrometer WASP to be realized at the ILL \cite{WASP}. However, the complexity of our set-up is significantly reduced \cite{prokudaylo}.

\section{Performance of RESPECT}

After having introduced the key features of RESPECT we show next how the optimization of the guide system was performed using Monte-Carlo simulations. Finally we introduce the expressions for the field integrals for NSE and LNSRE and demonstrate that  in the limit of vanishing divergence no field corrections are required for LNRSE in contrast to NSE, where field optimized coils are required. Therefore, LNRSE is particularly well suited for neutrons with a compact phase space.


\subsection{Brilliance of the cold pancake moderator of ESS}
\label{Mean_Brilliance}
\begin{figure}[htb]
\centering
\includegraphics[width=0.5\textwidth]{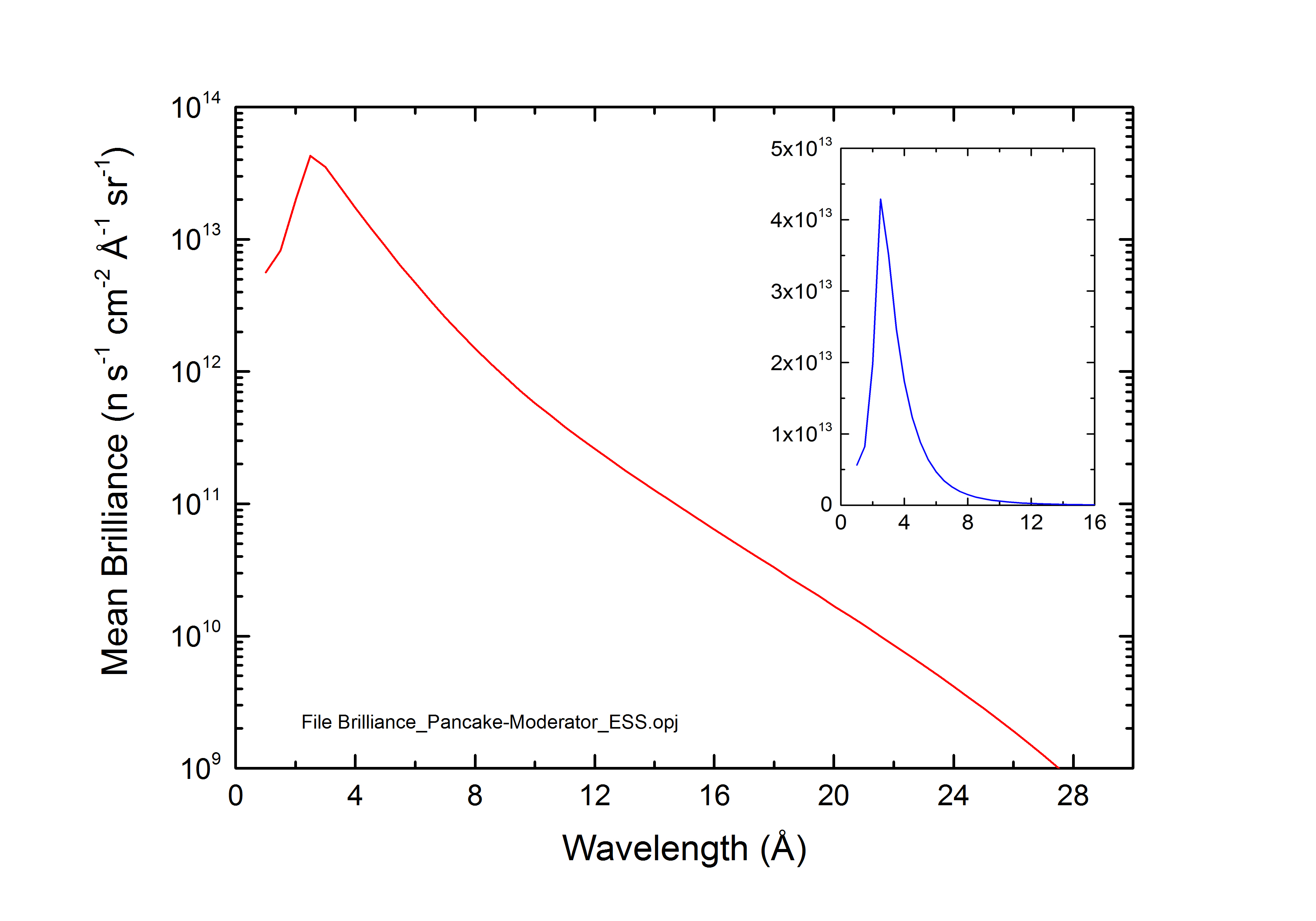}
\caption{Brilliance $\Psi$ versus $\lambda$ of the cold pancake moderator proposed for ESS \cite{ESS_mod_2}. $\Psi$ is largest around $\lambda = 3$ \AA, an important configuration for RESPECT if used for studies of single crystals.}
\label{Brilliance_Cold_Pancake} 
\end{figure}

The time-averaged brilliance of the pancake moderator of ESS \cite{ESS_mod_2}
is shown in Fig.~\ref{Brilliance_Cold_Pancake}. The moderator has the shape of a flat disc with a width of 320 mm and height of 30 mm.

Before embarquing into extensive Monte-Carlo simulations for the optimization of RESPECT we estimate the spectral flux density $F$ at the sample position of RESPECT for $\lambda = 6$ \AA\ using the expression
\begin{equation}
	F = \eta_{tot} \cdot \eta_{pol} \cdot \Delta\lambda \cdot \Omega \cdot \Psi
\label{flu}
\end{equation}
from reference \cite{Boeni:2014}. The parameters in Eq. (\ref{flu}) are assumed to be as follows:  From Fig.~\ref{Brilliance_Cold_Pancake}, the brilliance at 6 \AA\ is given by $\Psi = 4.7\cdot10^{12}$ s$^{-1}$cm$^{-2}$\AA$^{-1}$sr$^{-1}$. The solid angle at the sample is assumed to be $0.92^\circ \times 1.77^\circ = 4.96\cdot 10^{-4}$  rad (see Fig. \ref{Phase_Space_RESPECT}b). The wavelength band is $\triangle \lambda = 1$ \AA. The efficiency of the polarizer is $\eta_{pol} = 50\%$.

The transport efficiency $\eta_{tot}$ is of the order $\simeq 60\%$. It follows from the experience that the transmission of polarizing cavities for the correctly polarized neutrons is typically 80\% \cite{boeni2009} and the transmission of each bender is calculated to be 94\%. The efficiency of the beam extraction is estimated to be $\simeq 85\%$, i.e. not 100\% because the guide is vertically not fully illuminated.

The absorption of neutrons by Al, air, and He is neglected. For example, the transmission of 4 cm Al is 83\% for $\lambda = 6$ \AA\ . Inserting all parameters in Eq. \ref{flu} yields for the spectral flux density at the sample position $F(\lambda = 6\, {\rm \AA}) = 7.0\cdot10^8$ cm$^{-2}$s$^{-1}$\AA$^{-1}$. This value is indicated by a purple star in Fig. \ref{Intensity_PanCake} and is in excellent agreement with the Monte-Carlo simulations explained next.

\subsection{Monte-Carlo Simulations}
\label{BeamProperties}
\begin{figure}[htb]
\centering
\includegraphics[width=0.48\textwidth]{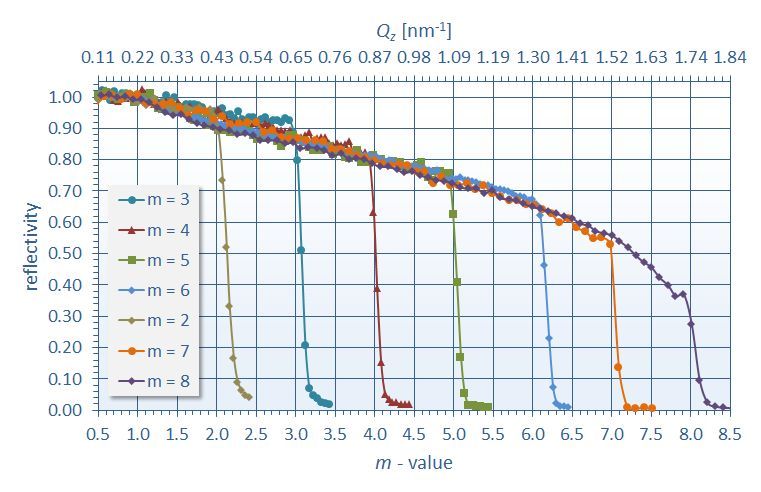}
\caption{The reflectivity of supermirrors decreases approximately linear with increasing index $m$, which is proportional to the momentum transfer $Q_z$ (upper scale of the figure) \cite{schanzer2016}.}
\label{Supermirror}
\end{figure}

Because of the small height of the pancake moderator, i.e. $h_{mod} = 30$ mm, the neutron guide must have a similar vertical dimension in order to take full advantage of the large brilliance of the moderator. If $h_{mod}$ is too large, the phase space of the neutrons becomes diluted thus reducing the brilliance downstream of the guide. Ideally, the entrance of the guide is moved as close as possible to the moderator, which is not possible at ESS, because a minimum distance of 2000 mm between the moderator and the entrance of the guide must be respected. For the reflectivity of the supermirror coatings we have used the experimental data shown in Fig. \ref{Supermirror}.

\begin{figure}[htb]
\centering
\includegraphics[width= 0.45\textwidth]{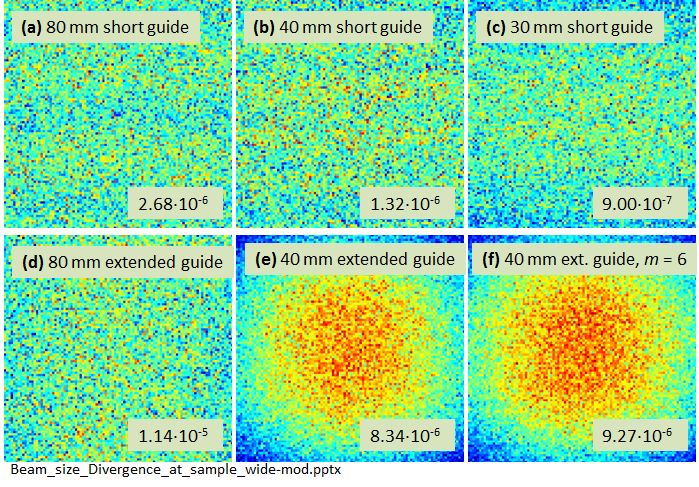}
\caption{Top and bottom row: The guide ends before and after the first precession region, respectively.  Each of the intensity patterns shows an area 30 mm $\times$ 30 mm. The beam size at the sample positions exceeds largely the dimensions of the sample if the neutron guide is large ((a), (d)) and/or ((a) - (c)) is not extended through the precession region. The proposed guide for RESPECT ($a = 40$ mm) confines the neutrons to the sample, which has a size of approximately 30 mm $\times$ 30 mm ((e), (f)). For the simulations, a moderator with a flat spectrum (component file: source\_gen.comp \cite{McStas}) of McStas was used. The numbers provide the integrated intensity over a beam area of 30 mm $\times$ 30 mm in units of s$^{-1}$.}
\label{Beam_size_at_sample_wide-mod}
\end{figure}

We have performed Monte-Carlo simulations using the software package McStas \cite{McStas} for optimizing the guide system shown in Fig. \ref{RESPECT_LNRSE}. First we demonstrate the influence of the cross section of the quadratic neutron guide with the dimensions $a \times a$ on the spectral flux density of the neutron beam at the sample position. Fig. \ref{Beam_size_at_sample_wide-mod} shows the expected result that terminating the neutron guide before the beginning of the first precession region leads to an over-illumination of the sample area exceeding significantly the anticipated sample size of 30 mm $\times$ 30 mm. For $a = 80$ mm there is an indication that the neutron spectral flux density is even inhomogeneous in the vertical direction showing two stripes of maximum spectral flux density which are due to the under-illumination of the guide by the moderator.

If a guide with $a = 40$ mm is chosen and extended throughout the first precession region, the sample is well illuminated. When compared with the configuration with $a = 80$ mm, the integrated spectral flux density is reduced by more than a factor of two due to the fact that the disk-shaped moderator could fully illuminate a guide with a width of 80 mm or more. The results also show that increasing the coating of the guide from $m = 2$ to $m = 6$ seems not to increase the integrated spectral flux density significantly and does not affect the beam profile visibly. However, as shown below, coatings $m = 6$ lead to an improvement of the transmission of neutrons with short wavelength.

\begin{figure}[htb]
\centering
\includegraphics[width= 0.4\textwidth]{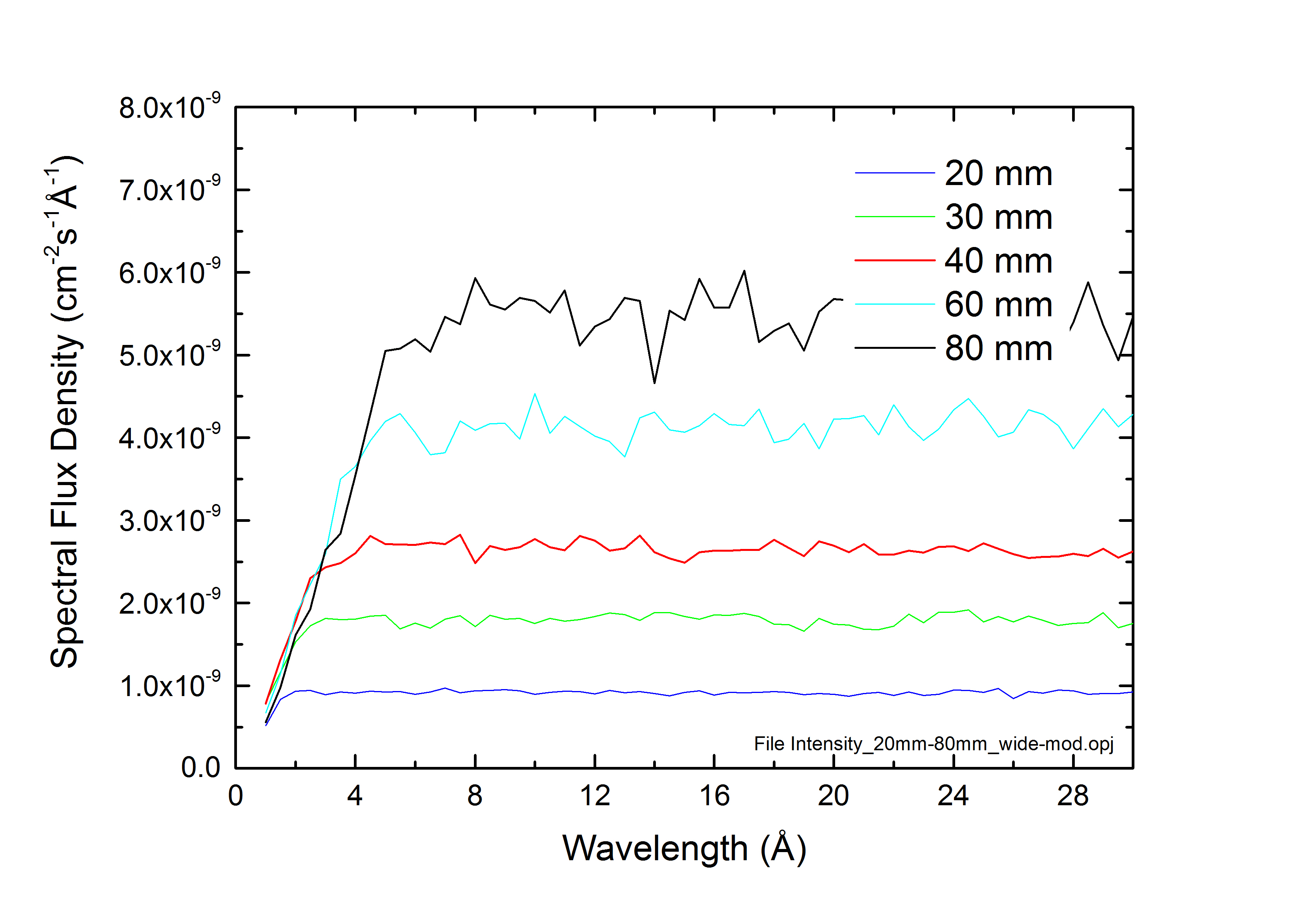}
\caption{The spectral flux density at the sample position (30 mm $\times$ 30 mm) is roughly proportional to the width $w$ of the guide because the wide moderator fully illuminates guides with widths 20 mm $\le w \le$ 80 mm.  \add{For the simulations, a moderator with a flat spectrum (component file: source\_gen.comp of McStas) was used.}}
\label{DependIntensCS}
\end{figure}

Fig. \ref{DependIntensCS}\ shows the expected result that the spectral flux density at the sample position of RESPECT increases approximately linearly (and not quadratically) with the dimension $a$ of the guide. The black and the red curves represent the results for the 80 mm and 40 mm guide, respectively. Note that guides with a large $a$ are less effective for short wavelengths. For these simulations the polarizer and the two bending devices were replaced by straight guide sections. The flat moderator of the software package McStas was used.

\begin{figure}[htb]
\centering
\includegraphics[width= 0.4\textwidth]{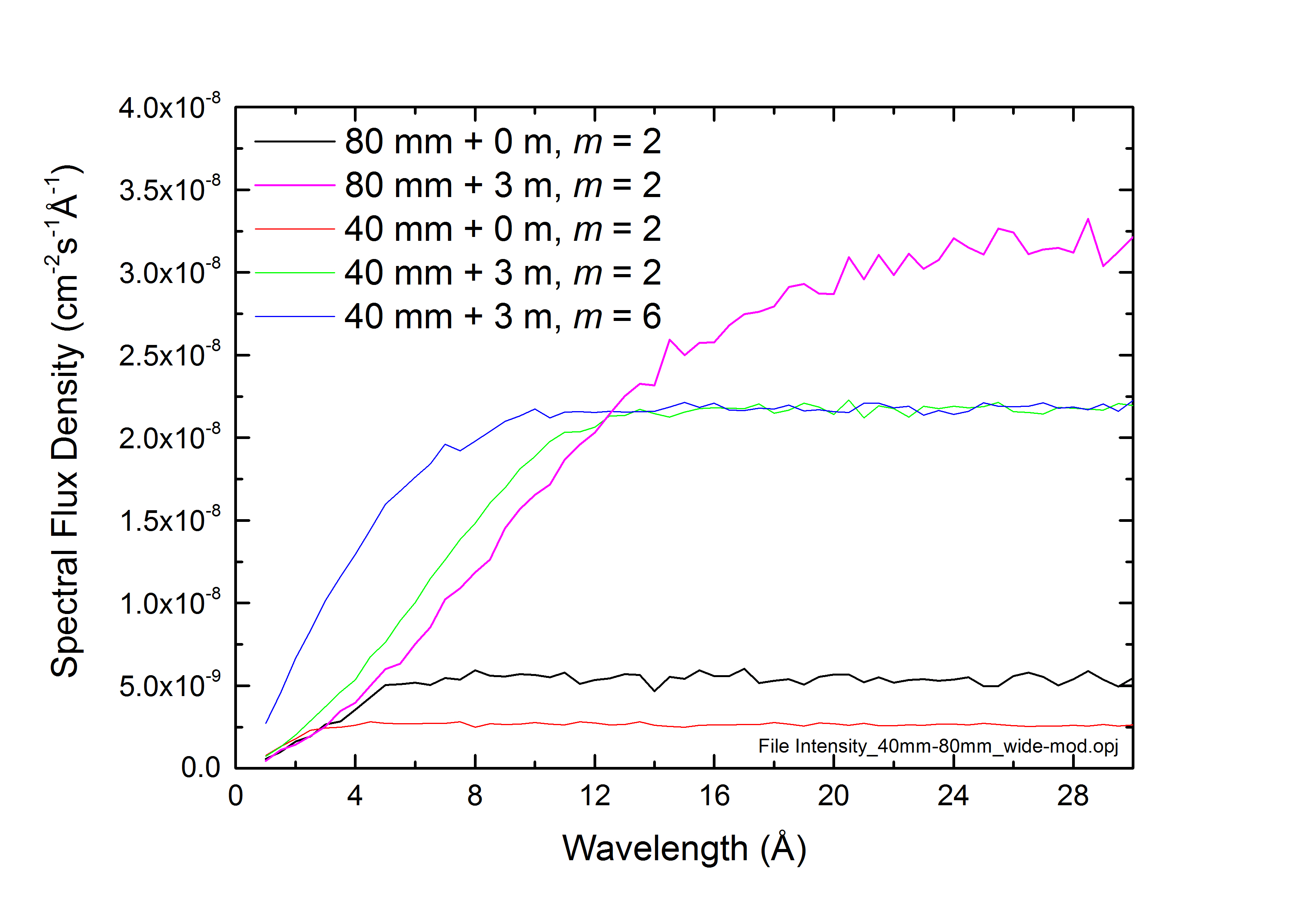}
\caption{Extending the neutron guide through-out the precession region leads to large gains in the spectral flux density. Coatings with $m = 6$ extend the critical wavelength of the guide system towards the regime of thermal neutrons. \add{For the simulations, a moderator with a flat spectrum (component file: source\_gen.comp of McStas) was used.}}
\label{DependIntensLength}
\end{figure}

Extending the guides through-out the precession region, i.e. by increasing their length by 3 m towards the sample leads to a rather dramatic increase in spectral flux density. As shown in Fig. \ref{DependIntensLength}, for $a = 40$ mm and 80 mm, the spectral flux density can be increased by about a factor of 6 and 4, respectively (see also the intensity values in Fig. \ref{Beam_size_at_sample_wide-mod}). Increasing the supermirror parameter $m$ from 2 to 6 leads to a lowering of the critical wavelength and may be considered as a valuable option for RESPECT to conduct experiments with thermal neutrons close to the maximum of the brilliance of the moderator ($\lambda \simeq 3$ \AA) more effectively.

\begin{figure}[htb]
\centering
\includegraphics[width=0.4\textwidth]{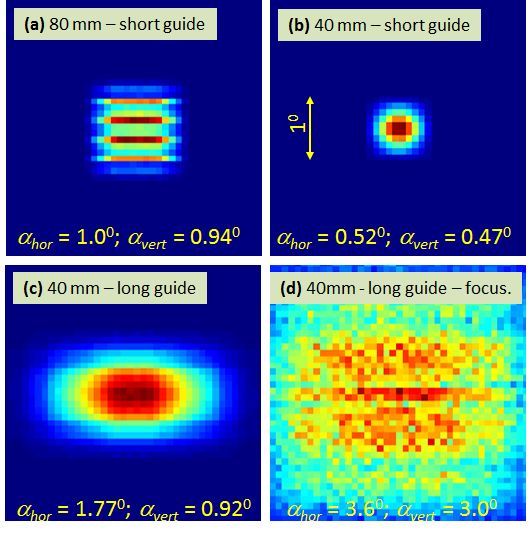}
\caption{(a), (b): The vertical divergence is very homogenous and inhomogenous for the 40 mm and 80 mm guide, respectively, while the horizontal divergence is homogeneous for both types of guides. (b): The 40 mm guide provides a very compact phase space at the sample, which is beneficial to achieve large $\tau$ in neutron spin echo. (c): By means of the additional 3 m long guide, the vertical and horizontal divergence can be increased by more than a factor of 2 and 3.5, respectively, leading to significant gains in intensity of approximately a factor of 4 -- 6 (see Fig. \ref{Beam_size_at_sample_wide-mod}). (d): Focusing the beam leads to a significant increase of the divergence.}
\label{Divergence_at_sample_wide-mod}
\end{figure}

Finally, we discuss the effect of $a$ on the divergence of the neutrons at the sample position. Fig. \ref{Divergence_at_sample_wide-mod} shows a comparison of the divergence for $a = 80$ mm and $a = 40$ mm as obtained for a sample size 30 mm $\times$ 30 mm. The divergence for the 40 mm guide is homogenous while for $a = 80$ mm, the vertical divergence becomes inhomogenous due to the under-illumination of the guide. Although the spectral flux density for the 80 mm guide is two times larger than for the 40 mm guide, the solid angle is even a factor of $\simeq 4$ larger, i.e. the phase space becomes diluted. 

Extending the 40 mm guide towards the sample leads to a significant increase of the divergence (and the intensity). The horizontal divergence is approximately 80\% larger for the extended 40 mm guide when compared with the 80 mm guide. These results show that the long guide with $a = 40$ mm is the preferred option for RESPECT. 

Using a parabolically focusing guide that starts after the first RF-coil and extending it to a distance $f = 200$ mm from the sample leads to an increase of the divergence by a factor of $\simeq 6$ which is compatible with the approximately six-fold increase of the intensity (red line in Fig. \ref{Intensity_PanCake}). Because the beams with the largest divergence are mostly focused after the second RF-coil, i.e. after the majority of the spin precessions is performed, it is expected that the maximum achievable $\tau$ will not deteriorate much. The experimental proof is pending.

\begin{figure}[htb]
\centering
\includegraphics[width=0.4\textwidth]{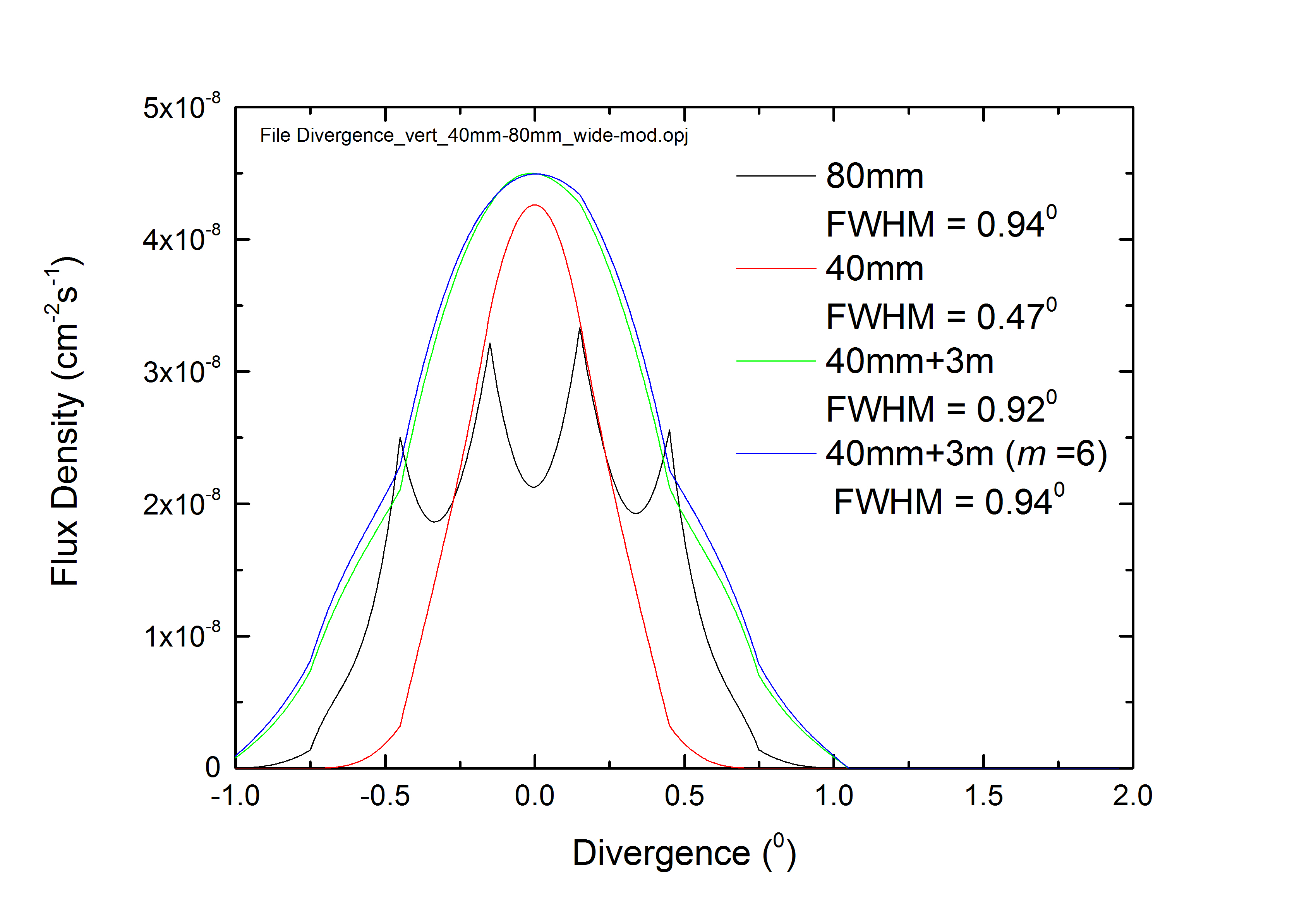}
\caption {The vertical divergence at the sample for a guide with $a =40$ mm is very homogeneous in contrast to $a = 80$ mm. Inserting a guide through-out the precession region (long guide) leads to an increase of the divergence at the sample. A supermirror coating $m = 6$ instead of $m = 2$ improves the transport of neutrons with short wavelengths (not shown). The divergence is not significantly affected by the increase of $m$.  \add{For the simulations, a moderator with a flat spectrum (component file: source\_gen.comp of McStas) was used.}}
\label{Divergence_vert}
\end{figure}

\begin{figure}[htb]
\centering
\includegraphics[width=0.4\textwidth]{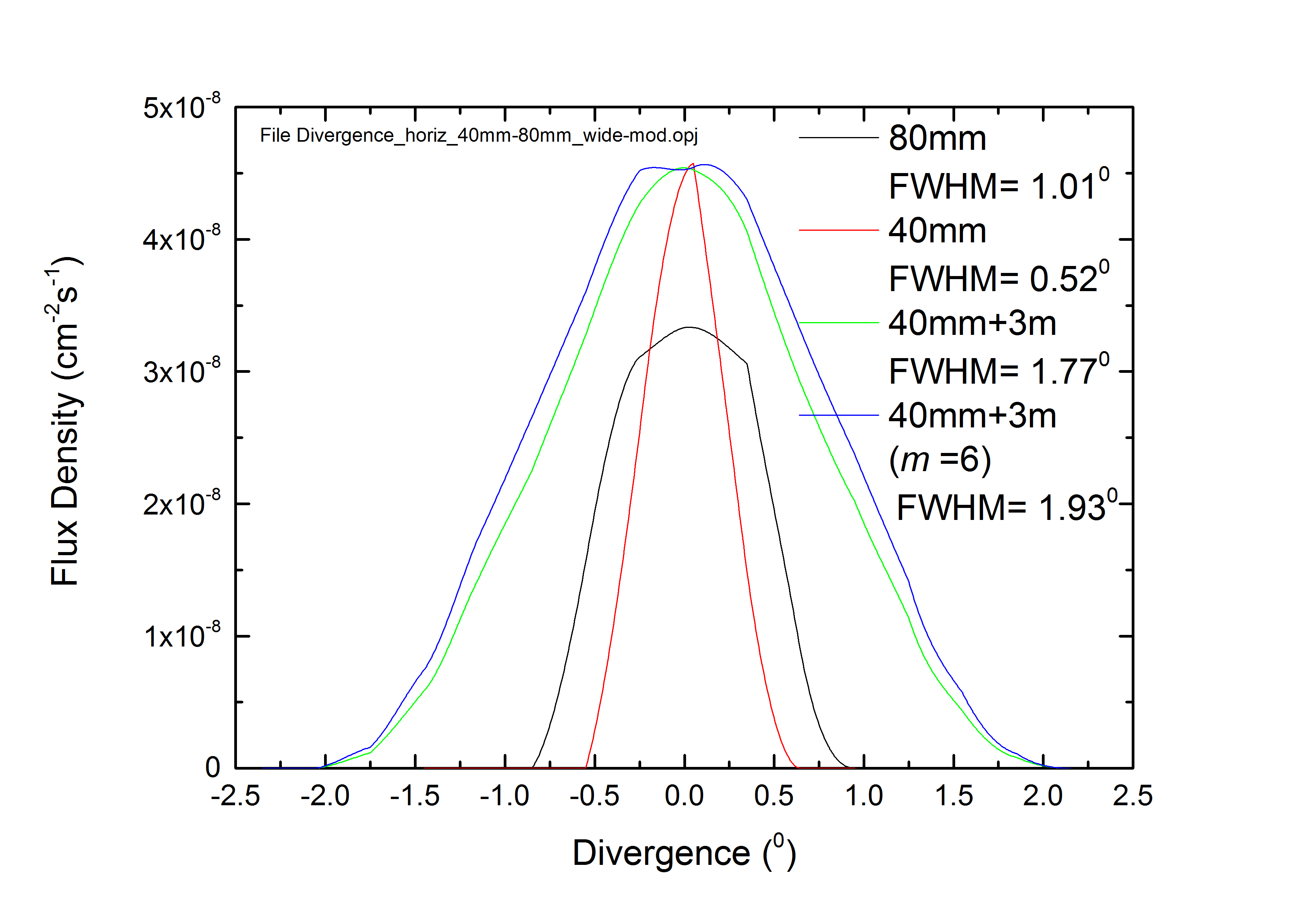}
\caption {The horizontal divergence is almost independent of $a$ for the same length of guide because the width of the moderator is very large $w_{mod} \gg 80$ mm. Extending the guide through the precession regions (long guide) leads to an increase in the horizontal divergence. A supermirror coating $m = 6$ instead of the proposed $m = 2$ improves the transport of neutrons with short wavelengths (not shown). The divergence is not significantly affected by the increase of $m$. \add{For the simulations, a moderator with a flat spectrum (component file: source\_gen.comp of McStas) was used.}}
\label{Divergence_horiz}
\end{figure}

A quantitative analysis of the divergence is shown in Figs. \ref{Divergence_vert} and \ref{Divergence_horiz}. Vertical and horizontal cuts through the contour plots (Fig. \ref{Divergence_at_sample_wide-mod}) are parametrized using Gaussians. The fitted divergencies are given in the figures. The results confirm the conclusions already reached.
The extension of the guide with $a = 40$ mm through the spin precession regions leads to an increase in the vertical divergence by a factor of two, i.e. to a similar value as for the 80 mm guide while the horizontal divergence is increased to $1.8^0$, i.e. by approximately 80\% when compared with $a = 80$ mm.

The concept of RESPECT forsees translating various collimations into the beam before the entrance of the Larmor precession region thus allowing to vary the horizontal and vertical divergence independently between 10 min and 60 min. Therefore, the divergence of the neutron beam can be reduced if required to reach very large spin-echo times. Of course, collimations $\alpha_{vert}$ and $\alpha_{horiz} < 1^0$ will partially eliminate the effects of the guide between the RF-coils on the divergence and intensity. If a parabolic focusing guide is inserted between the last RF-coil  and the sample, the insertion of the collimators allows tuning the size of the neutron beam at the sample.

\subsection{Homogeneity of field integrals}
\label{field integrals}

LNRSE as realized with the concept of RESPECT offers significant advantages to achieving long spin-echo times when compared with traditional NSE because the field corrections for non-divergent beams are essentially zero as shown below. Therefore, the adaption of the neutron guide of RESPECT to the pancake moderator  using a slim neutron guide with $a = 40$ mm leads to a neutron beam with a compact phase space, i.e. a small divergence (Fig. \ref{Divergence_at_sample_wide-mod}).

\begin{figure}[htb]
\centering
\includegraphics[width=0.45\textwidth]{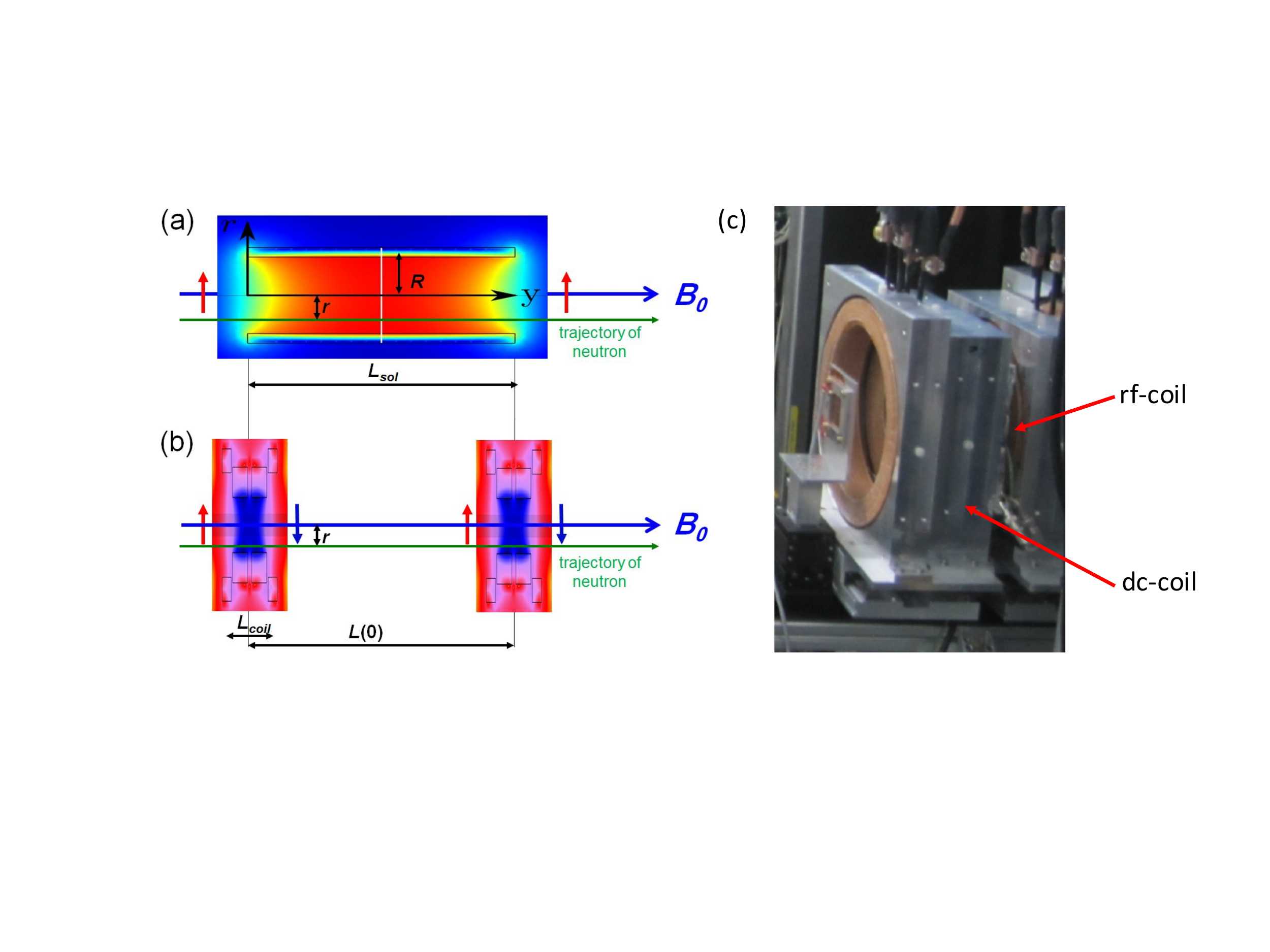}
\caption{Larmor precession regions for spin-echo spectrometers. (a) For NSE, a solenoid provides the static field. (b) In LNRSE, the precession region is confined between two RF-spin flippers. The $B_0$ fields are produced by coils in a Helmholtz configuration. Compensation coils reduce the stray fields at the field boundaries. The red and blue arrows indicate the spin directions of the neutrons entering and leaving the $B_0$ field regions. \add{(c) A longitudinal LRNSE coil at the instrument RESEDA. The rf-coil is placed between two dc-field coils.}}
\label{Field_Coils}
\end{figure}

Fig. \ref{Field_Coils} visualizes the field distributions of a solenoid as used for NSE (left hand side) and by a pair of partially compensated Helmholtz coils for LNRSE (right hand side), respectively. The colors indicate the distribution of the field amplitudes as calculated with the software tool COMSOL.

For a solenoid, the field integral $J(r)$ for neutrons with a divergence $\psi = 0$ is given by
\begin{equation}
  J(r) \simeq B_0 L_{sol} \biggl( 1 + {r^2 \over 2RL_{sol}} \biggr),
\end{equation}
where $L_{sol}$ and $R$ are the length and the radius of the solenoid, respectively, and $r$ the distance of the neutron trajectory from the axis of the solenoid. As the projections of the field gradients on the flight direction of the neutrons have the same sign at the entrance and exit of the solenoid, the inhomogeneities of the $B$-field add up and have to be corrected for, for example by means of Fresnel or Pythagoras coils \cite{MONKENBUSCH1990465}. 

In NRSE, the neutron spin is flipped by $\pi$ between the entrance and exit of each RF-spin flipper. Therefore, the effects of the field gradients on the neutron polarization cancel precisely for beams with zero divergence. Therefore, for properly manufactured and aligned Helmholtz-coils no corrections are required. The spin precession is performed in a homogeneous low field region between two RF-spin flippers. 

The effects of a finite divergence on the homogeneity of field integrals and the optimization of the field integrals in solenoids for NSE have been discussed in detail by Zeyen \cite{Zeyen1996}. Here, the concept of solenoids with an Optimal Field Shape (OFS) was introduced that was successfully applied to beam lines for NSE to correct for the finite divergence. 

\begin{figure}[htb]
\centering
\includegraphics[width=0.35\textwidth]{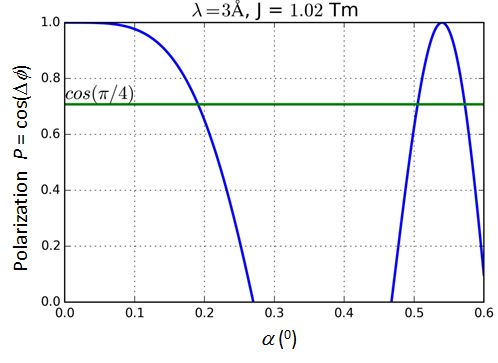}
\caption{Polarization $P$ versus the  inclination angle of the neutron beam. The field integral and neutron wavelength are $J = 1.02$ Tm and $\lambda = 3$ \AA, respectively (see Eqs. (\ref{Deltaphi},\ref{cosinusP}). The green line indicates the polarization for a phase difference $\Delta\phi = \pi/4$.}
\label{3A_94_Pol}
\end{figure}

For LNRSE, we estimate the influence of a finite divergence of the neutron beam on the precession as follows: The path length $L$, relevant for precession is given in the small angle approximation by
\begin{equation}
	L(\alpha) = {L(0) \over \cos\alpha} \simeq 1+ {\alpha^2 \over 2},
\label{path}
\end{equation}
where $L(0)$ is the path length for neutrons with $\alpha = 0$. $\alpha$ designates the inclination angle between the symmetry axis of the precession arm and the trajectory of the neutron. The neutrons attain a precession angle $\phi$ given by
\begin{equation}
	\phi = {\gamma_n m_n \over h}BL\lambda,
\label{precession}
\end{equation}
where $\gamma_n = 183.25 {\rm MHz / T}$, $m_n = 1.675\cdot10^{-27}$ kg, and $h = 6.626\cdot10^{-34}$ Js. By combining Eqs. (\ref{path}) and (\ref{precession}) one obtains for the phase difference of beams with an inclination $\pm \alpha$ when compared with beams along the axis ($\alpha = 0$)
\begin{equation}
	\Delta\phi = 4.632\cdot10^4 B {\rm [T]}\ L  {\rm [m]}\ \lambda {\rm [\AA]}\ {\alpha^2\over2}  {\rm [rad^2]}. 
\label{Deltaphi}
\end{equation}
If the polarization for a beam with $\alpha = 0$ is $P_i = 1$ then the polarization of an inclined beam is
\begin{equation}
	P = \cos(\Delta\phi).
\label{cosinusP}
\end{equation} 
$P$ versus the inclination angle $\alpha$ is depicted in Fig. \ref{3A_94_Pol} for a field integral $J = BL = 1.02$ Tm, i.e. the maximum field integral anticipated for RESPECT, and $\lambda = 3$ \AA. For example, for $\alpha = \pm 0.2^\circ$ one obtains $\Delta\phi = \pm 48^\circ$, which corresponds to a polarization $P(48^\circ) = 0.67$ (Fig. \ref{3A_94_Pol}).

For a uniform distribution of precession angles $\phi$ between $\pm \Delta\phi$ the average polarization is given by
\begin{equation}
	{P_{tot}} = \frac{1}{2\Delta\phi}  \int\limits_{-\Delta\phi}^{\Delta\phi} \cos(\Delta\phi) d\Delta\phi = {\sin \Delta\phi \over \Delta\phi}.
\label{ptot}
\end{equation}
For example, neutrons with a divergence $\psi = 2 \alpha = 0.4^\circ$ assume an average polarization of 89\% in agreement with the result quoted by Mezei \cite{1980:Mezei:book} p. 19.

We are now in a position to determine the maximum divergence $\psi$ of the beam that is allowed to achieve a minimum polarization $P$. It is assumed that the polarization of the incident neutrons is $P_i = 1$. If polarizations 
$P_{tot} = 90\%$ and $64\%$ are aimed for, according to Eq. (\ref{ptot}), the maximum allowed phase angles are $\Delta\phi = \pm 45^\circ$ and $\pm 90^\circ$, respectively. Fig. \ref{3A_94_Div} in section \ref{divergence-max} shows $\psi$ versus $\lambda$ using Eq. (\ref{Deltaphi}) for $P_{tot} = 90\%$ and 64\% assuming a field integral $J = 1.02$ Tm. It is seen that by restricting the divergence to $\psi = 0.2^\circ$, spin-echo measurements can be conducted at a polarization $P_{tot} = 64\%$ without using correction coils, i.e. spin echo times exceeding 1 $\mu$s can be reached for $\lambda = 20$ \AA\ (see section \ref{coilsystem}).

Of course, as in NSE it is possible to correct for the effects of divergent beams using Fresnel or Pythagoras coils, similarly as for non-divergent beams in NSE. According to Krautloher \cite{krautloher2014}, the field integral of a pair of RF-spin flipper coils with a separation $L(0)$ is approximately given by 
\begin{equation}
	J(r) = B_0 L(0) \biggl( 1 + {r^2 \over 2RL(0)} \cdot {L_{coil} \over L(0)} \biggr)
\end{equation}
where $L_{coil}$ is the length of the $B_0$ coil. When comparing this expression with the expression for NSE it is immediately seen that the power densities are much smaller because $L_{coil} \ll L(0)$.

\section{Conclusion}

We have demonstrated the concept of a high-resolution LNRSE-spectrometer using Larmor precessions based on the longitudinal spin-echo technique incorporating resonant RF spin-flipper coils. The energy and momentum resolution of RESPECT is equivalent to conventional state of the art neutron spin-echo spectrometers having the potential of reaching spin-echo times exceeding 1 $\mu$s. Using field subtraction coils between the RF spin-flipper coils the dynamic range of the instrument can be tuned over 8 orders of magnitude without need to change instrument components or instrument configurations. The open structure of the precession regions allows for the use of focusing neutron guides leading to potential gains of several orders of magnitude  when compared with existing NSE-spectrometers. A maximum spectral flux density exceeding $1\cdot 10^{10}$ cm$^{-2}$s$^{-1}$\AA$^{-1}$ polarized neutrons at the sample position will be feasible.

In the MIEZE-1 configuration, all spin manipulations are performed before the sample. Therefore, depolarizing samples such as ferromagnetic materials or materials containing hydrogen can efficiently be investigated. Moreover, magnetic fields can be applied in the sample region opening a flurry of new applications such as the study of field-driven quantum phase transitions and of vortex lattices in superconductors.  \change{Because no expensive magnetic shielding is required and standard components of reactor based neutron scattering instruments can be used, an LNRSE beamline can be installed in a cost-effective way.}{Because no expensive magnetic shielding is required and standard components of reactor based neutron scattering instruments (such as at RESEDA) can be used, an LNRSE beamline can be installed in a cost-effective way.}

\section{Acknowledgements}
This work was funded by  the German ministry for science and technologgy (BMBF) under ``Mit\-wirkung der Zentren der Helmholtz Gemeinschaft und der Technischen Uni\-ver\-sit\"{a}t M\"{u}n\-chen an der Design-Update Phase der ESS, F\"{o}r\-der\-kenn\-zeichen 05E10WO1'' \add{and also supported via the program "Erforschung kondensierter Materie an Gro\ss ger\"{a}ten". We also would like to thank the RESEDA team at the FRM II, Christian Franz, Thorsten Schr\"{o}der and Christian Fuchs. Georg Brandl, Kor\-binian Urban and Tobias Weber are acknowledged for their help with the McStas simulations.}

\bigskip
\bigskip

\noindent{\bf References}
\medskip
\bibliography{mieze-V9}

\end{document}